\documentclass{JHEP3}
\usepackage{latexsym}
\usepackage{epsfig}
\usepackage{epic}

\newcommand\cyr{%
\renewcommand\rmdefault{wncyr}%
\renewcommand\sfdefault{wncyss}%
\renewcommand\encodingdefault{OT2}%
\normalfont \selectfont} \DeclareTextFontCommand{\textcyr}{\cyr}

\title{Locality, Causality, and an Initial Value Formulation for
Open String Field Theory}

\author{Theodore G. Erler\\ University of California, Santa Barbara\\ Santa Barbara,
CA 93106, U.S.A\\ E-mail:\email{terler@physics.ucsb.edu}}

\author{David J. Gross\\ Kavli Institute of Theoretical Physics, Santa Barbara\\ Santa Barbara,
CA 93106, U.S.A\\ E-mail:\email{gross@kitp.ucsb.edu}}

\abstract{In this paper, we explore the questions of time,
locality and causality in the framework of covariant open bosonic
string field theory. We show that if an open string field is
expressed as a certain local function on spacetime---in
particular, a function of the lightcone component of the midpoint
and the transverse center of mass degrees of freedom---that cubic
string field theory is nonsingular and local in lightcone time. In
particular, the theory has a well defined initial value
formulation resembling that of an ordinary second order
relativistic field theory in lightcone frame. This description can
be achieved by a nonsingular unitary transformation on the Fock
space, and we demonstrate explicitly that the theory is gauge
invariant and the interaction vertex is local in this basis. With
an initial value formulation at hand, we are able to construct an
explicit second quantized operator formalism for the theory using
the Hamiltonian BRST formalism. We also explore issues of
causality by considering a singular limit of the theory where all
spacetime coordinates are taken to the midpoint. At any stage in
this limit, the theory is well-defined and arbitrarily close to
being completely local and manifestly causal. We argue that the
this limit must account for the macroscopic causality of the
string S-matrix.}


\preprint{hep-th/0406199}

\begin{document}
\def\a{\tilde{\alpha}}
\def\M{\mathrm{M}}
\def\P{\mathrm{P}}
\def\k{\kappa}
\def\EQN#1{eq.(\ref{eq:#1})}
\def\half{\!\matrix{\frac{1}{2}}}
\def\fraction#1#2{\matrix{\frac{#1}{#2}}}
\def\der#1{\frac{\partial}{\partial #1}}
\def\p2{\!\fraction{\pi}{2}\!}
\def\b{\tilde{b}}
\def\c{\tilde{c}}
\def\e{\epsilon}

\section{Introduction}
Causality is a basic requirement of any acceptable physical
theory; causes should always precede their effects, and if the
theory is Lorentz invariant, spacelike separated events should be
uncorrelated.

In quantum field theory, causality follows from a basic fact:
quantum field theory is {\it local}. In non-gravitational
theories, where the space-time manifold is fixed and
non-dynamical, causality means that the theory has localizable
observables which commute at spacelike separations, since
space-like separated measurements should not interfere. This
follows directly from locality, since observables are functions of
local fields that, as quantum operators, commute at space-like
separations (up to a gauge transformation) as a consequence of
Lorentz invariance and locality of the theory's interactions.
However, in many theories of interest the only known observables
are S-matrix elements describing scattering experiments, rather
than local observables. This is certainly true in gravitational
theories, where local operators are not diffeomorphism (gauge)
invariant, and it seems to be the case in string theory, even open
string theory. For such theories, causality requires certain
analytic properties of the S-matrix which ensure that, in a
scattering experiment, two incoming wave-packets will collide {\it
before} the outgoing wave-packets emerge. These analytic
properties also follow from locality: one expresses the scattering
amplitude in terms of Greens functions of interpolating fields,
either elementary or composite, which create asymptotic particle
states when acting on the vacuum. The locality of the commutators
of these interpolating fields leads to the analytic properties of
a causal S-matrix. This measure of macroscopic causality follows
from locality, but is weaker, since on-shell S-matrix amplitudes
are not enough to probe micro-causality. So it seems that locality
is an indispensable theoretical mechanism for ensuring that our
theories are sensible and causal.

\begin{figure}[top]
\begin{center}
\setlength{\unitlength}{1.5cm}
\begin{picture}(8,4)
\thicklines
\put(1,1){\line(2,1){1}} \put(2,1.5){\line(2,-1){1}}
\put(.95,1.05){\line(2,1){1}} \put(2.05,1.55){\line(2,-1){1}}
\put(2.05,1.55){\line(0,1){1}} \put(1.95,1.55){\line(0,1){1}}
\put(1,1){\line(0,-1){.5}} \put(3,1){\line(0,-1){.5}}
\put(.95,1.05){\line(-2,1){.5}} \put(1.95,2.55){\line(-2,1){.5}}
\put(3.05,1.05){\line(2,1){.5}} \put(2.05,2.55){\line(2,1){.5}}

\put(1.8,.65){\scalebox{1.5}[1.5]{$\Psi_1$}} \put(1,
1.75){\scalebox{1.5}[1.5]{$\Psi_2$}} \put(2.5,
1.75){\scalebox{1.5}[1.5]{$\Psi_3$}}

\put(1.1,.9){\scalebox{.75}[.75]{0}}
\put(1.95,1.3){\scalebox{.75}[.75]{$\p2$}}
\put(2.8,.9){\scalebox{.75}[.75]{$\pi$}}

\put(1.75,2.4){\scalebox{.75}[.75]{0}}
\put(1.75,1.65){\scalebox{.75}[.75]{$\p2$}}
\put(.9,1.15){\scalebox{.75}[.75]{$\pi$}}

\put(3,1.15){\scalebox{.75}[.75]{0}}
\put(2.1,1.65){\scalebox{.75}[.75]{$\p2$}}
\put(2.1,2.4){\scalebox{.75}[.75]{$\pi$}}

\put(5.5,1.5){\line(0,-1){1}} \put(5.5,1.5){\line(-1,1){1}}
\put(5.5,1.5){\line(1,1){1.3}}

\put(5.5,1.5){\line(4,1){1.5}}

\put(6.4, 2.475){\line(-1,1){.4}}

\put(7.0,1.875){\line(0,-1){1.2}} \put(7.0,1.875){\line(1,1){1}}

\dottedline[.]{.07}(7.0,1.875)(6.5,2.375)

\put(5.35,1.4){\scalebox{.75}[.75]{0}}
\put(7.05,1.775){\scalebox{.75}[.75]{$\pi$}}

\put(6.0,.9){\scalebox{1.5}[1.5]{$\Psi_1$}} \put(5.4,
2.3){\scalebox{1.5}[1.5]{$\Psi_2$}} \put(6.8,
2.3){\scalebox{1.5}[1.5]{$\Psi_3$}}

\put(.3,.5){a)}\put(4.7,.5){b)}

\end{picture}
\end{center}
Figure 1: a) The Witten vertex $\langle \Psi_1,
\Psi_2*\Psi_3\rangle$. b) A hypothetical, but obviously incorrect,
``local'' interaction of strings.
\end{figure}

How does this discussion extend to string theory? At first sight
it would appear that strings, being extended objects, could not be
described by a local, casual theory. It seems natural to address
this question directly in the framework of covariant string field
theory. In the case of open bosonic strings we have a particularly
simple formulation, due to Witten\cite{Witten}, in terms of a
cubic spacetime action, \begin{equation} S = -\langle \Psi,
Q_B\Psi\rangle -\frac{2}{3}g\langle \Psi, \Psi*\Psi\rangle
\label{Witten_action}\end{equation} where $\langle,\rangle$ is the
BPZ inner product, $Q_B$ is the BRST operator corresponding to a
choice of conformal background, and the string field $\Psi$ is a
vector in the state space of a particular boundary conformal field
theory (see \cite{Taylor-Review} for a nice review). The cubic
interaction term is given by the Witten vertex, which can be
described as follows: Writing the string field as a functional of
an open string configuration $\Psi[x(\sigma)]=\langle
x(\sigma)|\Psi\rangle,\ \ \sigma\in[0,\pi]$ (ignoring ghosts), to
calculate $\langle \Psi_1, \Psi_2*\Psi_3\rangle$ we must glue the
left half of $x(\sigma)$ in $\Psi_1$ to the right half of
$x(\sigma)$ in $\Psi_2$ and so on around cyclically, as shown in
figure 1a. Certainly, this procedure is very nonlocal. By
comparison, a naive ``local'' interaction $\Psi[x(\sigma)]^3$
(figure 1b) clearly has nothing to do with string interactions as
we know them.

Perhaps then it comes as a surprise that critical string theory
produces an analytic S-matrix consistent with macroscopic
causality. In absence of any other known theoretical mechanism
which might explain this, despite appearances one is lead to
believe that string interactions must be, in some sense, local.

The problems with nonlocality and causality in string theory
become almost fatal when we realize that the theory appears to be
nonlocal in time. To see the problems more deeply, suppose we
truncate the cubic action eq.\ref{Witten_action} to include only
the tachyon field $\Psi=\int dp\phi(p)|p\rangle$:
\begin{equation}
S_{\mathrm{tachyon}} = \int dx \left[\phi(\half\partial^2+1)\phi -
\frac{2}{3}\kappa g [e^{\frac{1}{2}
V_{00}\partial^{2}}\phi]^{3}\right]\label{tach}
\end{equation}
Due to the nasty differential operator $e^{\frac{1}{2}
V_{00}\partial^{2}}$ this action contains an infinite number of
derivatives in both time and space (if we included more fields we
would get an action with similar, but more complicated
appearance). The Lagrangian eq.\ref{tach} is completely nonlocal,
as can be seen with the formula\footnote{Actually, there is even
some subtlety in defining the differential operator
$e^{\frac{1}{2} V_{00}\partial^{2}}$\cite{Hata-Time}. Our formula
gives a definition which converges only for a certain class of
functions.},
$$e^{\frac{1}{2} V_{00}{\partial}^2}\phi(x)=\frac{1}{2\pi V_{00}}\int dx'
e^{\frac{1}{2 V_{00}} (x-x')^2}\phi(x')$$ Apparently, the tachyon
at a point $x$ couples not only to itself, but also to its values
arbitrarily far in the future, in the past, and even at space-like
separations! Such a theory could not be meaningfully causal. Nor
could one imagine that it has a sensible initial value
formulation. Without an initial value formulation, we cannot
proceed to the Hamiltonian formalism and canonically quantize, so
it is not clear that a quantum theory for the action eq.\ref{tach}
even exists. Perhaps the most frightening aspect of eq.\ref{tach},
however, is that any theory whose Lagrangian depends nontrivially
on any more than first time derivatives has a completely unstable
Hamiltonian\cite{Ostrogradski,Time-Problem}. The instabilities
presumably present in eq.\ref{tach}, however, could not manifest
themselves in the perturbative S-matrix, since the higher
derivatives enter only at the level of the interaction. Certainly,
we hope that string theory has a causal, stable Hamiltonian and a
sensible quantum mechanical definition beyond perturbation theory;
yet it has been a mystery how string theory really manages to
escape these sicknesses.

It is therefore clear that any acceptable physical theory should
be local in time. If the theory is Lorentz invariant, presumably
this means that it is (in some sense) local in space as well, and
this should be sufficient to ensure macroscopic causality of the
S-matrix.

In this paper we attempt to face up to these facts in the
framework of open bosonic string field theory. We find that string
theory avoids problems with nonlocality in a surprising way. In
particular, we find that the Witten vertex is ``local enough'' to
allow for a nonsingular description of the theory which is
completely local along a single null direction. Specifically, if
we regard the string field as a certain function on spacetime---a
function of the lightcone component of the midpoint and the
transverse center of mass degrees of freedom---the cubic action is
local and first order in lightcone time derivatives. Therefore
string field theory has a well-defined initial value formulation,
a sensible canonical quantum theory, and a Hamiltonian free of
higher derivative instabilities. However, since our choice of
spacetime coordinate is not Lorentz invariant, microscopic
causality is not manifest and the theory remains nonlocal in the
transverse spatial directions. The remaining nonlocality, we find,
cannot be removed while maintaining a nonsingular description of
the theory. This perhaps should be expected, since the known
ultraviolet properties of string theory, in particular ultraviolet
finiteness and polynomial boundedness of the S-matrix, seem to
contradict properties of a completely local quantum field theory.
The picture we find is therefore similar to that in lightcone
string field theory: the theory is local in (lightcone) time but
still nonlocal in space. However, unlike lightcone string field
theory, it is clear that cubic string field theory at least has a
local limit where all spacetime coordinates are taken to the
midpoint. We investigate this limit with a careful choice of
regulator and show that at any stage the theory is nonsingular but
arbitrarily close to being local and manifestly causal. We believe
that the existence of this limit, though singular, must account
for the macroscopic causality of the string S-matrix. Thus, string
theory is local enough to avoid the inconsistencies of a theory
which is acausal and nonlocal in time, but is nonlocal enough to
make string theory different from quantum field theory.

To motivate our particular perspective, it seems appropriate to
discuss earlier attempts to understand the role of locality,
causality and time in string theory, and explain why we feel these
approaches do not adequately address the problems just raised. To
start with, we mention some discussions of causality in the
framework of free lightcone\cite{Martinec} and
covariant\cite{Hata-Causal} string field theory. It has been
argued that the natural generalization of the commutativity of
quantum fields at space-like separations is that string fields
should commute outside the so-called {\it string light cone},
\begin{equation}
\int d \sigma \bigl[ x(\sigma)-x'(\sigma) \bigr]^2 >0\ \
\rightarrow\ \
\big[\Psi\bigl(x(\sigma)\bigr),\Psi\bigl(x'(\sigma)\bigr)] =0
\label{str_lc}
\end{equation}
This is a very strange condition. It is not reparameterization
invariant, and although correct for free strings, is is violated
once interactions are included\cite{Susskind}.

It must be said that the meaning and necessity of eq.\ref{str_lc}
is far from clear. String fields are not observable nor is it
clear how to construct observables from them; there is no reason
why their commutators should have any locality properties at all.
What is required for the establishment of macroscopic causality is
the analytic properties of the scattering amplitudes, which is
related directly to locality properties of the correlation
functions. Correlation functions of string fields expressed as
functionals of $x(\sigma)$ correspond to path integrals over
Riemann surfaces with holes and half-disks removed; such objects
are highly singular and ill-suited for constructing scattering
amplitudes for asymptotic particle states in string theory. To
construct a more appropriate basis of interpolating fields one
should decompose the string field into a mode basis of ordinary
local fields, $\phi^i(x)$, one for each observable asymptotic
particle described by the string. These can serve as a basis of
interpolating fields for the purpose of constructing the S-matrix,
and if they satisfy causal commutation relations
$$[\phi^i(x),\phi^j(y)]=0\ \ \ \ (x-y)^2>0 $$
(though perhaps in a singular limit) then macrocauslity should be
valid. At any rate, for free strings, the Lagrangian is simply a
sum of free Lagrangians for the individual string modes, and
trivially the component fields commute at space-like separation.
Causality is only an issue once interactions are included.
Therefore we feel that eq.\ref{str_lc} is not a good starting
point for discussing issues of locality and causality in string
theory.

Apparently, string field theory should not be regarded as
specifying an action for some singular functional
$\Psi[x(\sigma)]$, but rather as specifying a action for a
countable number of local spacetime fields $\phi^i(x)$, one for
each mode of the string. To resolve the difficulties of
nonlocality and causality in string theory, we need to show that
there is some choice of $\phi^i(x)$ which at the very least
renders the theory local in time, and perhaps in some singular
limit local in space as well. A natural question then arises: what
is the label $x$ in all of these fields? In quantum field theory,
the meaning of $x$ is clear: it refers to the location $\vec{x}$
of a point particle at time $x^0$. In our case, however, the
string is not a point particle; whatever $x$ describes must depend
on how we choose to ``break up'' the string into particle-like
constituents. While we do not have a unique notion of ``position''
in string theory, we do have a well-defined notion of momentum:
the conserved charge $p_\mu$ associated with translations of
$x(\sigma)$. It is natural to require that an acceptable choice of
$x$ satisfies,
$$[x^\mu, p_\nu]=i\delta^\mu_\nu.$$ This condition
follows for any $x$ given by,
\begin{equation}x^\mu=\int_0^\pi d\sigma f^\mu_\nu(\sigma)x^\nu(\sigma)
\ \ \ \ \ \int_0^\pi d\sigma f^\mu_\nu(\sigma)=\delta^\mu_\nu\ \ \
\ \label{x}\end{equation}  Thus, having decided that local fields
are what interest us, we must still decide, subject to eq.\ref{x},
what our fields are local in.

In most studies of string field theory to date, the standard
choice for $x$ has been the string center of mass
$x_{\mathrm{cm}}=\frac{1}{\pi}\int d\sigma x(\sigma)$. The center
of mass has the advantage of being the natural spacetime
coordinate for the mass eigenstates of the free string. However,
the disadvantage of $x_\mathrm{cm}$ is that the action appears
extremely nonlocal in both space and time, as we saw in
eq.\ref{tach}.

The problems with the center of mass $x$ have recently become
unavoidable in the context of the tachyon condensation problem in
cubic string field theory. Let us recall the basic story. After
the seminal work of Sen\cite{Sen1} it has been realized that the
tachyon of the open bosonic string can be interpreted as an
instability of the space-filling D-25 brane on which the open
string ends. If the brane is allowed to decay, one is presumably
left with a vacuum without any D-branes or open strings, i.e. the
vacuum of the closed bosonic string. Strong evidence for the
validity of this conjecture comes from the level truncation
scheme, where one truncates the cubic action to include only a
finite number of lightest mass fields. At zero momentum, the
Lagrangian is reduced to a quadratic plus cubic polynomial
potential of scalar fields whose minimum is an approximation to
the closed string vacuum. Remarkably, level truncation converges
rapidly\cite{Tachyon}, and yields an excellent approximation to
the known (or conjectured) exact result for the difference in
vacuum energies between the unstable and stable solutions.

It is natural given a potential of this form to study time
dependent solutions, starting close to the unstable maximum and
rolling down to the stable vacuum. Sen in fact proposed a boundary
conformal field theory describing such a process, whereby the
tachyon rolls homogenously towards the closed string vacuum but
does not cross over in finite time\cite{Sen}. Attempts to identify
such a solution using level truncation of the mass eigenstates in
open string field theory, however, have run into serious problems.
Solutions seem to have very erratic behavior, in complete
qualitative disagreement with Sen's rolling tachyon solution: the
string field seems to pass quickly through the closed string
vacuum and then quite far up the steep side of the potential,
after which a sequence of oscillations of diverging amplitude
ensues\cite{Hata-Time,Zwiebach-Time}. In retrospect, the
pathological runaway behavior of these solutions is not much of a
surprise, given the higher derivative instabilities expected from
the extreme nonlocality of the cubic action when truncated in mass
eigenstates\footnote{For some additional studies of the nonlocal properties of truncated open string field theory and p-adic string theory, see ref.\cite{More} }.

To restore sanity one is tempted to consider the lightcone string
field theory, which although gauge fixed and not manifestly
Lorentz invariant, has the advantage of being ghost free and
completely local in lightcone time $x^+$. The difficulty with this
approach, however, is that we have no concrete evidence that
lightcone string field theory contains nonperturbative
information, and in particular no stable closed string vacuum
solution is known. In fact, some\cite{Time-Problem} have argued
that that the theory's locality in $x^+$ is solely an artifact of
perturbation theory. Indeed it is not difficult to see that there
is something of a paradox with the theory's locality in $x^+$: the
lightcone string field theory only assigns one physical phase
space degree of freedom per component field, whereas the covariant
theory, containing an infinite number of time derivatives, assigns
an infinite number. The authors of ref.\cite{Time-Problem} offered
a possible resolution to this paradox in terms of the mechanism of
``localization,'' which can be understood as follows. In studying
a theory whose higher time derivatives enter only in interaction,
one can always identify two types of solutions: 
perturbative solutions, which pass over in the weak coupling limit
to solutions of the free theory, and nonperturbative solutions  which do
not. The nonperturbative solutions generically carry negative energies and display unstable, ``runaway'' behavior. For such theories, it turns out to be possible to find an
``equivalent'' theory without higher derivatives whose solution
space contains only the perturbative ``stable'' solutions, but not
the runaway ones\footnote{To see how localization
works\cite{Time-Problem}, consider a simple model of a particle at
position $q$ satisfying the equation of motion,
$$0=\left(\frac{d^2}{dt^2}+ \omega^2\right)q +
\frac{g(1-g)}{\omega^2}\frac{d^4}{dt^4}q$$ In this model, the
quartic time derivative is the ``interaction,'' whose strength is
measured in $g$. It is simple to see that equation has four
linearly independent solutions, two positive energy solutions with
frequencies $\pm\frac{\omega}{\sqrt{1-g}}$ and two negative energy
solutions with frequencies $\pm \frac{\omega}{\sqrt{g}}$. Only the
positive energy solutions are well defined in the $g\to 0$ limit,
and they can be described just as well with the second order
equation of motion,
$$0=\left(\frac{d^2}{dt^2}+ \frac{\omega^2}{1+g}\right)q$$ Thus,
perturbatively the theory is equivalent to one without higher
derivatives. For many more generic and complicated examples, we
refer the reader to ref.\cite{Time-Problem}.}. Possibly this is
how lightcone string field theory achieves its locality in $x^+$.
If this were true, then nonperturbatively lightcone string field
theory would be inequivalent to covariant string field theory, and
in particular probably fails to be Lorentz invariant. One can hope that this is not the
case, but clearly it would be much more reassuring to establish
some sort of locality (particularly in $x^+$) directly in
covariant string field theory.

Thus, other approaches having proved inadequate, we are forced to
return to covariant string field theory and consider whether we
can do better by choosing a position coordinate {\it other} than
the string center of mass to label the spacetime dependence of our
component fields. Keep in mind that, as far as the closed string
vacuum is concerned, any choice of $x$ is equally valid, since the
vacuum is translationally invariant and hence independent of $x$.

A little thought about the Witten vertex reveals that the only
choice of $x$ which has a hope of giving us a local action for the
component fields is the string midpoint, $x(\p2)$. Indeed, there
were several papers in the early days\cite{Morris,Manes,Potting}
which attempted to formulate string field theory directly in terms
of fields which were local functions of the midpoint.
Unfortunately such a formulation appears to be extremely singular:
the component fields all carry infinite energy and the vertex is
generically afflicted with anomalies which spoil locality. The
central insight of this paper, however, is that we do not need a
completely local and nonsingular formulation in terms of component
fields; all we need is locality in a single direction which can be
identified as time. This is possible: if we choose $x$ so that its
light cone component $x^+$ coincides with the string midpoint, the
component fields can have finite energy and the vertex is
manifestly local in lightcone time.

This paper is organized as follows. In section 2 we explain why it
is not possible to have a completely local formulation of string
field theory in terms of the midpoint coordinate. Specifically,
the kinetic terms in the action are infinite and the vertex
generally fails to be local, contrary to naive expectations. In
section 3 we identify a much less singular basis (the ``tilde
basis'') where only lightcone time $x^{+}$ is identified with the
string midpoint, and explain how this can be achieved by a
nonsingular unitary transformation on open string the Fock space.
We also introduce a convenient basis for the ghost fields where
the zero mode of $c(\sigma)$ is taken to be $c(\p2)$. With the
help of the tilde basis we elucidate the initial value formulation
and identify constraints which must be imposed on the initial
conditions. In section 4 we construct an explicit second-quantized
operator formalism for the interacting string field theory. We
identify a remarkably simple spacetime BRST charge and prove that
it is both nilpotent and commutes with the string Hamiltonian. As
an added bonus, we show that our choice of basis in the ghost
sector allows one to see explicitly that the classical master
action for string field theory automatically satisfies the {\it
quantum} BV master equation, thus providing a field theoretic
proof that the Feynmann diagrams of the cubic action automatically
provide a complete single cover of the moduli space of open
Riemann surfaces. In section 5 we explore the issue of causality,
in particular whether we can show given our initial value
formulation that information propagates only in the future light
cone. Since lightcone time plays a preferred role in the tilde
basis, causality is not manifest. However we explore the singular
limit where all components of $x$ are taken to the string
midpoint. With a careful choice of regulator, we show that at any
stage in the limit the theory is nonsingular and arbitrarily close
to being manifestly Lorentz invariant and local, and hence
presumably causal. In appendix A we carefully study the
interaction vertex in the tilde basis, and demonstrate both
analytically and numerically that the theory is both gauge
invariant and completely local in lightcone midpoint time for
well-behaved string fields. We have placed this discussion in an
appendix since it is somewhat technical, but it is crucial for
establishing the credibility of our results---experience shows
that string field theory is a delicate framework which easily
disintegrates if arguments are overly formal. In appendix B we
prove some identities used in appendix A, and in appendix C we
list some useful formulas. In section 6 we offer some conclusions.

After much of this work was completed, we realized that the basic
features of the lightcone midpoint formalism had already been
discovered many years ago by Maeno\cite{Maeno}, though his work
seems to be completely unknown. We feel that these ideas are
sufficiently fundamental to be brought again to the attention of
the community. At any rate, our work goes substantially beyond
Maeno's in providing an modern and detailed discussion, an
explicit second quantized operator formalism for the
theory\footnote{By contrast, Maeno constructs the Hamiltonian path
path integral for the theory, related in a fairly straightforward
way to the usual Lagrangian path integral of Thorn\cite{Thorn}.},
some exploration of the role of causality, and a careful
demonstration of both gauge invariance and locality.

\section{Why the midpoint doesn't work}
Before moving on to the body of our paper, it seems necessary to
explain why {\it only} the lightcone component of the midpoint can
be understood as defining a time coordinate in which cubic string
field theory is local and no more than second order in time
derivatives. Indeed, since the cubic interaction identifies the
midpoint coordinates of all three string fields locally, one would
imagine that any timelike component of the midpoint could be used
to construct a well-defined initial value formulation of the
theory. In fact, one might even propose that the string field
should be viewed as a spacetime function of all components of
midpoint, in which case the theory should be completely local and
second order in the midpoint coordinate. This idea was originally
proposed by Witten\cite{Witten-Super} and was subsequently
explored in references \cite{Morris,Manes,Potting}. Recently the
idea has reappeared in the context of the operator/Moyal formalism
in ref.\cite{Belov,Bars}. Our discussion follows that of
references \cite{Morris,Manes,Potting,Belov}. The idea is to
implement a unitary transformation, \begin{equation}U =
\exp\left[-p_\mu \sum_{n=1}^\infty
\frac{(-1)^n}{2n}(\alpha^\mu_{2n} - \alpha^\mu_{-2n})\right].
\label{U_wrong}\end{equation} on the mode basis of the state space
of the boundary conformal field theory. Under this transformation
the string center of mass $x$ becomes the string midpoint position
$x(\p2)$:
$$Ux^\mu U^{-1}=x^\mu(\p2).$$ The tachyon vacuum state $|k\rangle$
transforms to a new vacuum state $|k\rangle'$ labelled by momenta
$k$ which are now interpreted as conjugate to the midpoint
position, rather than the string center of mass. The spacetime
fields corresponding to the modes in this basis should satisfy
field equations whose kinetic term is second order and whose
quadratic nonlinear term contains no derivatives.

The trouble is that the unitary operator eq.\ref{U_wrong} is
singular, as can be seen by viewing its normal ordered form, $$U =
\exp\left[-\half p^2\sum_{n=1}^\infty
\frac{1}{2n}\right]\exp\left[p_\mu \sum_{n=1}^\infty
\frac{(-1)^n}{2n}\alpha^\mu_{-2n}\right]\exp\left[-p_\mu
\sum_{n=1}^\infty\frac{(-1)^n}{2n}\alpha^\mu_{2n}\right] $$ The
first factor here involves a divergent sum. Even more troubling is
the appearance of $L_0$ in this basis, \begin{equation}L_0 = \half
p^2+\sum_{n=1}^\infty \bar{\alpha}_{-n}\cdot\bar{\alpha}_n + p_\mu
\sum_{n=1}^\infty(-1)^n(\bar{\alpha}^\mu_{2n}
+\bar{\alpha}^\mu_{-2n})+p^2\sum_{n=1}^\infty 1
\label{L_0wrong}\end{equation} where $\bar{\alpha}=U\alpha
U^{-1}$. The last term is infinite, so it seems that $L_0$ is
undefined in this basis. We interpret this divergence as meaning
that the fields in this basis, though (naively) satisfying local
field equations, have infinite energy and are unphysical. In fact,
due to the singular nature of $U$ locality of the interaction is
even a subtle issue. One approach\cite{Morris} to regulating these
divergences is to replace $(-1)^n/2n$ in eq.\ref{U_wrong} with
$\omega_{2n}/2n$ with $\omega_{2n}=\lambda^n(-1)^n$ and take the
limit $\lambda\to 1^-$. In this regularization the vertex turns
out {\it not} to be local in the limit $\lambda\to 1^-$ contrary
to reasonable expectations.

Another approach explored in ref.\cite{Potting} involves
$\zeta$-function regularization, where one sets $\omega_{2n} =
(2n)^{-s}(-1)^n$ and takes the limit $s\to 0^+$. In this approach
they were able to demonstrate the locality of the vertex; moreover
the divergent sum in $L_0$ is now reinterpreted as $-\half$, by
the magic of $\zeta$ function regularization. However, though the
theory appears nonsingular and local in this regularization, it
fails to be gauge invariant at $s=0$ due to ambiguities in
defining $Q_B^2$; moreover the physical perturbative states are
undefined in this basis due to the singular nature of $U$ at
$s=0$. While these troubles can be avoided at nonzero $s$, the
theory is only local at $s=0$ so it seems that we still do not
have a well defined local formulation of the theory. In our
opinion the $\zeta$-function regulator is probably not sensible
anyway, since we do not believe it is correct make the analytic
continuation $\sum 1\to-\half$. The kinetic term $\langle
\Psi|c_0(L_0+L_0^{gh}-1)|\Psi\rangle$ is not an analytic function
of $s$ since for complex $s$ the transformation generated by $U$
on the Fock space is not unitary, so clearly the kinetic term for
a particular component field will depend both on $s$ and $s^*$. At
any rate, the divergence of $L_0$ in the midpoint basis probably
has a physical origin and should not be argued away.

One point should be kept in mind when discussing these regulators:
at any stage, the uncertainty in the expectation value of $x(\p2)$
remains infinite. To see this, consider the transformed tachyon
state at a point $x$:
$$|x,\omega\rangle =U_\omega|x\rangle$$
where $\omega$ refers to $\omega_{2n}$, our chosen midpoint
regulator. The ``uncertainty'' in the expectation value of
$x(\p2)$ can be defined in terms of the root mean square
deviation,
$$\langle \Delta x(\p2)\rangle^2(\omega) \equiv
\frac{1}{\delta(0)}\langle x,\omega| x(\p2)^2 |x,\omega\rangle -
\left[\frac{1}{\delta(0)}\langle x, \omega|
x(\p2)|x,\omega\rangle\right]^2$$ Calculating this we find,
$$\langle \Delta
x(\p2)\rangle^2(\omega)=D\sum_{n=1}^\infty\frac{((-1)^n-\omega_{2n})^2}{2n}$$
Unless the midpoint limit has already been reached, for any
acceptable regulator $\lim_{n\to\infty}\omega_{2n}=0$. Therefore,
this sum is logarithmically divergent. It would therefore seem
dubious that these regulators actually ``approach'' the
midpoint\footnote{In the midpoint lightcone basis, the cross terms
which generate this divergence are absent. Therefore, at least in
a light-like direction, we can meaningfully converge to the
midpoint}.

\section{Lightcone basis}
Let us now describe explicitly the choice of basis which renders
the theory local in lightcone time. Consider a spacetime vector
$v^\mu = (v^0, v^1, v^2,...)$. Define lightcone components,
\begin{equation}v^+ \equiv -v_- =\fraction{1}{\sqrt{2}}(v^0+v^1)\ \ \ \
v^- \equiv -v_+ =\fraction{1}{\sqrt{2}}(v^0-v^1)\end{equation} It
is useful to introduce two vectors $\lambda$ and $\chi$
satisfying,
\begin{eqnarray}\lambda_\mu\lambda^\mu &=& \chi_\mu\chi^\mu = 0\
\ \ \lambda_\mu\chi^\mu = -1 \nonumber\\  \lambda_\mu v^\mu &=&
v^{+}\ \ \ \ \ \ \ \ \ \ \ \ \ \chi_\mu v^\mu =
v^{-}.\end{eqnarray} In addition define
\begin{eqnarray}v^\mu_\M &\equiv& (\delta^\mu_\nu +
\lambda^\mu \chi_\nu)v^\nu =
(\fraction{1}{\sqrt{2}}v^+,\fraction{1}{\sqrt{2}}v^+,v_2,...)\nonumber\\
v^\mu_\P &\equiv& (\delta^\mu_\nu + \chi^\mu \lambda_\nu)v^\nu =
(-\fraction{1}{\sqrt{2}}v^-,\fraction{1}{\sqrt{2}}v^-,v_2,...)\nonumber\\
v^\mu_\perp &\equiv& (\delta^\mu_\nu + \lambda^\mu \chi_\nu +
\chi^\mu\lambda_\nu)v^\nu = (0,0,v_2,...).\end{eqnarray} $v_\M$
denotes $v$ with it's minus component set to zero, $v_\P$ is $v$
with its plus component set to zero, and $v_\perp$ is $v$ with
both its plus and minus components set to zero.

Consider the state space $\mathcal{H}_\mathrm{BCFT}$ of the
boundary conformal field theory describing an open bosonic string
living on a space-filling D25 brane. The usual basis for
$\mathcal{H}_\mathrm{BCFT}$ is given by the mode oscillators
$\alpha^\mu_n, b_n, c_n$ acting on the vacuum $|k\rangle$
describing the open string tachyon at momentum $k$ ($\alpha^\mu_0
= p^\mu$). We consider a change of basis generated by the unitary
transformation,
\begin{equation} U = \exp\left[-p_+ \sum_{n=1}^\infty
\frac{(-1)^n}{2n}(\alpha^+_{2n} - \alpha^+_{-2n})\right].
\label{U}\end{equation} Under this change of basis the matter
oscillators and zero-modes transform as,
\begin{eqnarray}\a^\mu_n &\equiv& U\alpha^\mu_n U^{-1} = \alpha^\mu_n -
 \cos\fraction{n\pi}{2} p_+ \lambda^\mu
\ \ \ \ \ n\neq 0 \nonumber\\ \tilde{p}_\mu
 &\equiv& U p_\mu U^{-1}= p_\mu \nonumber\\ \tilde{x}^\mu &\equiv& U x^\mu U^{-1}
 = x^\mu -i\sqrt{2}\chi^\mu \sum_{n=1}^\infty\frac{(-1)^n}{2n}(\alpha^+_{2n}
 - \alpha^+_{-2n}). \label{tilde_basis}\end{eqnarray}
The ghosts are unaffected. In particular, the plus component of
$\tilde{x}$ lies at the string midpoint while the other components
lie on the center of mass:
\begin{equation} \tilde{x}^+ = x^+(\fraction{\pi}{2})\ \ \ \ \
\tilde{x}_\P^\mu = x_\P^\mu \label{zero_modes} \end{equation}
Naively, then, we expect string field theory to be local and first
order in $\partial_+$. The vacuum $|k\rangle$ transforms into a
state $|k\rangle'$:
\begin{equation}|k\rangle' \equiv \exp\left[k_+\sum_{n=1}^\infty
\frac{(-1)^n}{2n}\alpha^+_{-2n}\right]|k\rangle,\label{vacuum1}\end{equation}
Since the transformation is unitary, this basis satisfies the
usual properties,
\begin{eqnarray}[\a^\mu_m, \a^\nu_{-n}] &=&
m\eta^{\mu\nu}\delta_{mn}\ \ \ \ \ [b_m,c_{-n}] =\delta_{mn} \nonumber \\
 \a^\mu_n|k\rangle' &=&
b_n|k\rangle' = c_n|k\rangle' = 0\ \ \ \ \ \ \ n>0\nonumber\\
p_\mu|k\rangle' &=& k_\mu|k\rangle'\ \ \ \ b_0|k\rangle' = 0
\end{eqnarray} The zeroth Virasoro generator takes the form:
\begin{eqnarray}L_0 &=& \tilde{L}_0|_0 + \half p^2 + p_+
\sum_{n=1}^\infty(-1)^n(\a_{2n}^+ +\a_{-2n}^+)\nonumber\\
&=& \tilde{L}_0|_0 + \half p_\perp^2 +
p_+P^+(\p2)\label{L_0}\end{eqnarray} $P(\p2)$ is the momentum of
the string midpoint (see appendix C for a definition of $P(\p2)$
and other midpoint coordinates in following equations). Our
notation is that a tilde over an operator denotes that operator
with the replacement $\alpha\rightarrow\a$, and $|_0$ means we set
the zero modes to vanish: $p=b_0=c_0=0$. Though $L_0$ is
nondiagonal in this basis, it is finite and well defined. This
means, in particular, that states created by acting a finite
number of $\a,b,$ and $c$s on the new vacuum $|k\rangle'$ have
finite energy. Hence, we seem to have a nonsingular and physical
basis for describing string fields with the crucial advantage that
the theory in this basis should appear local and first order in
lightcone time derivatives.

At this point one might make an objection to our approach: by
isolating one particular lightcone component of the position and
translating it to the midpoint, we have spoiled manifest Lorentz
invariance. This is really not the case, since cubic string field
theory is Lorentz invariant and all we have done is chosen a
particular basis for describing it. What we really mean when we
say ``manifest Lorentz invariance'' is that Lorentz
transformations are generated by {\it linear} transformations of
the fields in the theory. This is as true in our basis as in any
basis. In particular, the generator of Lorentz transformations in
the old basis is, \begin{equation}J^{\mu\nu} = x^{[\mu}p^{\nu]} +
i\sum_{n=1}^\infty \frac{1}{n} \alpha^{[\mu}_{-n}\alpha^{\nu]}_n
.\end{equation} In the tilde basis it is:
\begin{equation}J^{\mu\nu}=\tilde{J}^{\mu\nu} +
\chi^{[\mu}p^{\nu]}\bar{X}^+ +
p_+\lambda^{[\mu}\bar{X}^{\nu]}\end{equation} where
\begin{equation}\bar{X}^\mu = i\sum_{n=1}^\infty\frac{(-1)^n}{2n}(\a^\mu_{2n}
-\a^\mu_{-2n}) = x^\mu(\p2)-x^\mu.\end{equation} What is different
about the tilde basis is that Lorentz transformations not only
transform the position coordinates and the spacetime indices, as
in usual field theory, but also transform between different
spacetime fields corresponding to different modes in the basis.
However, it is important to realize that if one were to try to
approximate the theory by truncating fields beyond some level in
the tilde excitations, Lorentz symmetry is lost. In the old basis,
Lorentz invariance is preserved at any order in the level
truncation scheme.

Let us see how some important operators in the theory appear in
the tilde basis. The Virasoro operators are:
\begin{eqnarray}L_{2n}&=&\tilde{L}_{2n}|_0 + p_\M\cdot\a_{2n}
+ \pi(-1)^np_+P^+(\p2)\nonumber\\L_{2n-1}&=&\tilde{L}_{2n-1}|_0 +
p_\M\cdot\a_{2n-1} -i(-1)^np_+x^+(\p2)'\end{eqnarray} In writing
the BRST operator in the tilde basis, it is useful first to
separate explicitly its dependence on zero modes:
\begin{equation}Q_B = c_0(\half p^2+L_0|_0+L_0^{gh}-1)- b_0A +
p\cdot B +Q_B|_0\label{BRST_old}\end{equation} where,
\begin{eqnarray}A &=&
2\sum_{n=1}^\infty nc_{-n}c_n \nonumber\\ B^\mu &=&
\sum_{n=1}^\infty(c_n\alpha^\mu_{-n} + c_{-n}\alpha^\mu_n)
\end{eqnarray} The BRST operator in the tilde basis can then be
written, \begin{eqnarray}Q_B&=&\pi c(\p2)p_+P^+(\p2) - i\pi
p_+x^+(\p2)'\pi_b(\p2)\nonumber\\ &\ &+c_0(\half p_\perp^2
+\tilde{L}_0|_0+L_0^{gh} -1) - b_0A + p_M\cdot\tilde{B} +
\tilde{Q}_B|_0\label{BRST}\end{eqnarray} Of particular importance
is the first term of this equation. Our interest in this term
stems from the fact that it can be interpreted as responsible for
the dynamics of the string field. It is the only term in the
string field theory action where $p_+$ appears multiplied by
$p_-$, and Fourier transforming $p_+p_- = -\partial_+\partial_- =
\half(\partial_0^2 -\partial_1^2)$ which contains the familiar
second time derivative generating time evolution. The thing to
notice about this term is that it is multiplied by $c(\p2)$ while
the corresponding $p_+p_-$ term in the old basis is multiplied by
$c_0$, as can be seen by inspecting eq.\ref{BRST_old}. The
relevance of this fact is as follows. The string field can be
written as the sum of two terms: $$|\Psi\rangle = |\phi\rangle +
|\psi\rangle$$ where $b_0|\phi\rangle=0$ and $c_0|\psi\rangle=0$.
The familiar choice of Siegel gauge corresponds to setting
$|\psi\rangle=0$. The kinetic term in the string field theory
action can be written in terms of $|\phi\rangle$ and
$|\psi\rangle$ \begin{eqnarray} S &=&
-\langle \Psi,Q_B\Psi\rangle\nonumber\\
&=& -\langle \phi, c_0(\half p^2 + L_0|_0 + L_0^{gh}-1)\phi\rangle
-\langle \psi, b_0A\psi\rangle - 2\langle \psi,(p\cdot
B+Q_B|_0)\phi\rangle \nonumber\end{eqnarray} Note that, because
$c_0$ annihilates $|\psi\rangle$, second time derivatives only
appear acting on $|\phi\rangle$, and so $|\phi\rangle$ is the only
component of the string field which is truly dynamical. From a
Hamiltonian perspective $|\psi\rangle$ represents gauge degrees of
freedom since its conjugate momentum vanishes identically.
However, from eq.\ref{BRST} we can see that with respect to the
midpoint lightcone time $|\psi\rangle$ is no longer non-dynamical,
since $c(\p2)$ does not annihilate $|\psi\rangle$. This suggests
that, to separate the dynamical and gauge degrees of freedom, it
is more useful to decompose the string field as, $$|\Psi\rangle =
|\phi\rangle' + |\psi\rangle' $$ where $b_0|\phi\rangle'=0$ and
$c(\p2)|\psi\rangle'=0$.

Therefore it seems useful to perform yet another change of basis
on the ghost sector so that the zero mode of the $c$ ghost is
$c(\p2)$ rather than $c_0$. This change of basis can be
implemented by the unitary transformation,
\begin{equation}U_{gh}\equiv
\exp\left[-b_0\sum_{n=1}^\infty(-1)^n(c_{2n}+c_{-2n})\right]
\label{U_gh}\end{equation} The ghost oscillators transform as,
\begin{eqnarray}\b_n &\equiv& U_{gh}b_n U^{-1}_{gh}
= b_n - b_0\cos\fraction{n\pi}{2} \ \ \ \ \ n\neq 0\nonumber\\
\c_n &\equiv& U_{gh}c_n U^{-1}_{gh} = c_n \ \ \ \ \ n\neq 0 \nonumber\\
\b_0 &\equiv& U_{gh}b_0 U^{-1}_{gh} = b_0 \nonumber \\
\c_0 &\equiv& U_{gh}c_0 U^{-1}_{gh} = c_0 +
\sum_{n=1}^\infty(-1)^n(c_{2n}+c_{-2n}) =
c(\p2)\label{ghost_tilde}\end{eqnarray} The vacua transform as
\begin{equation}|-,k\rangle' \equiv |k\rangle' = U_{gh}|k\rangle'\
\ \ \ \ \ \ |+,k\rangle' \equiv U_{gh}c_0|k\rangle'
\end{equation} In particular, the $|k\rangle'$ vacuum does not
transform (we will usually suppress the $-$ when denoting this
vacuum). Since the transformation is unitary, the basis satisfies
the expected properties, \begin{eqnarray}&\
&[\b_m,\c_{-n}]=\delta_{mn} \nonumber\\
&\ & \b_n|k\rangle' = \c_n|k\rangle' = 0\ \ \ \
n>0\nonumber\\
&\ & \b_n|+,k\rangle' = \c_n|+,k\rangle' = 0\ \ \ \
n>0\nonumber\\
&\ & b_0|k\rangle'=0, \ \ \ \ \c_0|+,k\rangle'=0 \end{eqnarray}
The ghost Virasoros take the form, \begin{eqnarray}L_0^{gh} &=&
-\pi b_0\pi_b(\p2)'+\tilde{L}_0^{gh}|_0\nonumber\\
L_{2n}^{gh} &=& 2n\b_{2n}(\c_0-\bar{C})-\pi(-1)^n
b_0\pi_b(\p2)'+\tilde{L_{2n}^{gh}}|_0\nonumber\\
L_{2n-1}^{gh} &=& (2n-1)\b_{2n-1}(\c_0-\bar{C}) - (2n-1)\pi(-1)^n
b_0\pi_b(\p2)\nonumber\\
&\ &\ \ \ \ + (-1)^n b_0 c(\p2)'+\tilde{L}_{2n-1}^{gh}|_0
\end{eqnarray} where, $$\bar{C}\equiv
\sum_{n=1}^\infty(-1)^n(\c_{2n}+\c_{-2n})=c(\p2)-c_0$$ The BRST
operator takes the somewhat complicated form,
\begin{eqnarray}Q_B &=& \pi p_+\left[\c_0 P^+(\p2) -ix^+(\p2)'\pi_b(\p2)\right] +
(\c_0-\bar{C})\left[\half p_\perp^2 +\tilde{L}_0|_0\right]\nonumber \\
&\ &
+\c_0\tilde{L}_0^{gh}-\half(\bar{C}\tilde{L}_0^{gh}+\tilde{L}_0^{gh}\bar{C})
+\p2\pi_b(\p2)'[\c_0\b_0-\b_0\c_0]\nonumber\\ &\ &
 + p_\M\cdot\tilde{B} +
\tilde{Q}_B|_0\label{BRST2}\end{eqnarray} Note that now when we
write a tilde over an operator, this means we replace all
oscillators with their tilde'd counterparts in both matter and
ghost sectors. Also $|_0$ means we set the zero modes to vanish in
the tilde basis. Remember however that in eq.\ref{BRST} the tilde
denotes replacing {\it only} the oscillators in the matter sector.
From here on, when we talk about the tilde basis we mean making
the unitary transformations eq.\ref{U} and eq.\ref{U_gh} in both
matter and ghost sectors.

The BPZ inner product in the tilde basis takes exactly the same
form as it does in the old basis. In particular, we have the
familiar relations, \begin{eqnarray}
\langle\a_{-m}^\mu\Psi,\Phi\rangle &=& (-1)^{m+1}\langle
\Psi,\a_m^\mu\Phi\rangle\nonumber\\
\langle\b_{-m}\Psi,\Phi\rangle &=& (-1)^{m}(-1)^{\Psi}\langle
\Psi,\b_m\Phi\rangle\nonumber\\
\langle\c_{-m}\Psi,\Phi\rangle &=& (-1)^{m+1}(-1)^{\Psi}\langle
\Psi,\c_m\Phi\rangle
\end{eqnarray} where $(-1)^\Psi$ denotes the Grassmann parity of $\Psi$.
The two string vertex $\langle V_2|$ is as in the old basis after
the replacement of the oscillators and vacua with their tilded
counterparts.

It is worth noting that all numerical calculations in string field
theory performed in Siegel gauge with $p_+=0$ translate directly
in our formalism, since in this context the tilde basis is
identical to the old basis. This includes the vast majority of
successful calculations in the theory performed to date. For
calculations outside of Siegel gauge, such as those calculating
the spectrum of fluctuations around various D-brane vacua, or for
calculations of time independent spatially inhomogeneous solutions
which happen to depend on $x^1$, our basis will of course yield
different results. It is worth exploring the level truncation
scheme in the tilde basis to see whether it represents a usefully
convergent approximation to the theory. As mentioned earlier, the
only explicitly time dependent solution produced in the level
expansion sofar seems to display strange behavior due to the
infinite number of time derivatives, and it is not altogether
clear whether the solution should be taken seriously. Our hope is
that the tilde basis provides a better framework for studying time
dependent solutions in the theory. Investigations along these
lines are currently under way and will be published in
ref.\cite{me}.

We expect that in the tilde basis the matter part of the three
string vertex will be local in lightcone time. What should happen
to the ghost part of the vertex? The ghost
vertex\cite{Gross-Jevicki} satisfies the following overlap
condition:
\begin{equation}c^{(A)}(\sigma)|V_3\rangle =
-c^{(A+1)}(\pi-\sigma)|V_3\rangle\ \ \ \ \
\sigma\in[0,\p2]\end{equation} where the index $A=1,2,3,
\mathrm{mod}3$ denotes the Hilbert space on which the $c$ ghost
acts. Suppose that this equation holds strictly even in the
boundary case $\sigma=\p2$. The vertex should then satisfy:
\begin{eqnarray}c^{(1)}(\p2)|V_3\rangle &=&-c^{(2)}(\p2)|V_3\rangle
=c^{(3)}(\p2)|V_3\rangle\nonumber\\&=&-c^{(1)}(\p2)|V_3\rangle
\nonumber\end{eqnarray} This implies the surprising result,
$$c^{A}(\p2)|V_3\rangle = 0$$ This means, at least naively,
that any string field $|\psi\rangle'$ satisfying
$\c_0|\psi\rangle'=0$ will have vanishing star product with any
other field:\begin{equation}\psi'*A=0\label{vertex0}\end{equation}
In fact, in an earlier paper Okuyama\cite{Okuyama1} considered a
basis for the the ghost state space equivalent to our tilde basis,
and he showed that fields annihilated by $\c_0$ do in fact satisfy
this property. For completeness in the next section we study this
property explicitly, together with the locality of the vertex, and
attempt to understand the circumstances under which it can be
expected to hold. An immediate consequence of eq.\ref{vertex0} is
that the star product should have no identity element. This is
surprising since there has long been known a string field
$|I\rangle$ which seems to behave as an identity element when
multiplied with well-behaved string fields. Apparently, the field
$|\psi\rangle'$ is not well behaved by this criterion. In fact,
the expression for the identity string field involves explicitly a
midpoint insertion of the $c$ ghost momentum:
\begin{equation}\pi_c(\p2)=\frac{1}{\pi}\left[b_0 +
\sum_{n=1}^\infty(-1)^n(b_{2n}+b_{-2n})\right]\label{pic}\end{equation}
A brief look at eq.\ref{ghost_tilde} reveals that this operator is
undefined in the tilde basis, as must be identity string field.
Similar problems with the identity are encountered in vacuum
string field theory\cite{VSFT} where the kinetic operator
$\mathcal{Q}=c(\p2)$ has no well defined action on $|I\rangle$.
Fortunately, though in early formulations $|I\rangle$ was used to
define the integral $\int$ in the string field theory action, the
identity turns out to be unnecessary in the fundamental
formulation of the theory--- the old integral is now replaced by
the BPZ inner product $\langle,\rangle$ which is completely
nonsingular in the tilde basis. Still, one might be bothered by
the fact that our choice of basis has rendered such an important
element of the algebra singular. However, we emphasize that in
order to give string field theory a well defined initial value
formulation it is not necessary to transform the ghosts, and for a
particular application one may choose not to for the sake of
salvaging the identity string field\footnote{Analogously, in the
matter sector the midpoint lightcone $+$ component of the momentum
$P_+(\p2)$ is singular in the tilde basis. Fortunately, we are not
aware of any important operators or fields in string field theory
where $P_+(\p2)$ enters in a crucial way.}. Still, even if we do
not transform the ghosts, we would need to regulate expressions
involving both the identity string field and the BRST operator
since the BRST operator in the tilde basis depends explicitly on
$c(\p2)$.

At this point we should explain precisely in what sense the tilde
basis allows us to define the initial value problem for open
string field theory. Since locality in time can only be achieved
in the lightcone frame and the theory possesses gauge invariance,
the initial value problem is somewhat more complicated than in
nondegenerate second order systems where time evolution is
uniquely determined by specifying the coordinates and velocities
at $t=0$. First off, to solve for the evolution of the string
field uniquely we must fix a gauge. A natural choice is Siegel
gauge, which although afflicted with Gribov problems\cite{Taylor}
is sufficient for our discussion. The Siegel gauge equations of
motion are,
\begin{equation}\left[\partial_+\partial_- -\partial_\perp^2
-i\partial_+ \sum_{n=1}^\infty(-1)^n(\a_{2n}^+ +\a_{-2n}^+)
+\tilde{L}_0|_0 +L_0^{gh}-1\right]\Psi(x) +
gb_0\Psi*\Psi(x)=0\label{Siegel_EOM}\end{equation} To see how this
equation of motion appears for a typical component field in the
tilde basis, it is useful to consider a simplified model with all
the relevant features. Consider a $1+1$ dimensional field theory
for two scalars $\Phi(x,t)=(\phi(x,t),B(x,t))$ (writing $t=x^+$
and $x=x^-$ to avoid index clutter) with the equation of motion,
\begin{equation} (-\partial_t\partial_x
+1)\phi(x,t)+\partial_t B(x,t) +\int_{-\infty}^\infty
dydz\Phi(y,t)\cdot G(x,y,z)\Phi(z,t)=0\label{model}\end{equation}
where $G$ is some coupling matrix depending on three copies of
$x$. We interpret this as an equation of motion for $\phi$ which
is first order in $t=x^+$ but due to the interaction is not local
in $x=x^-$. The fact that the evolution of $\phi$ is coupled the
time derivative of $B$ corresponds to the $-i\partial_+
\sum_{n=1}^\infty(-1)^n(\a_{2n}^+ +\a_{-2n}^+) $ term in
eq.\ref{Siegel_EOM}, which couples each component field in the
tilde basis to the time derivatives of an infinite number of other
fields which differ from it by one fewer $\a^+_{2n}$ excitation
and one greater $\a^-_{2n}$ excitation. To determine the evolution
of $\phi$ we integrate this equation with respect to
$x^-$:\begin{equation} \dot{\phi}(x,t)=\dot{\phi}(c,t) +\int_c^x
dw\left[\phi(w,t)+\dot{B}(w,t) +\int_{-\infty}^\infty
dydz\Phi(y,t)\cdot G(w,y,z)\Phi(z,t)\right]\nonumber\end{equation}
Let us assume for the moment that $B$ and its time derivative are
known functions. From this equation, we can see that if we specify
the initial configuration of $\phi$ on the lightlike surface $t=0$
and the value of $\dot{\phi}$ on the orthogonal lightlike surface
$x=c$, we can determine the time evolution of the field uniquely.
Note that the nonlocality of the interaction in $x=x^-$ plays no
role in this statement. However, one might be skeptical that the
term ``initial value formulation'' really applies to this system,
since we not only need information about the field at $t=0$, but
at all times on the surface $x=c$. This is true, but it is worth
noting that $c$ can be chosen arbitrarily; in particular, we can
choose $c=-\infty$ corresponding to early times and large spatial
distances. For physically reasonable solutions we expect $\phi$ to
vanish out there, so $\dot{\phi}(-\infty,t)=0$ and the only
remaining boundary condition is $\phi$ at $t=0$, which is indeed
an ``initial'' condition.

Of course this prescription assumes that $\dot{B}$ is already
known. In practice this is tricky, since to solve for the time
derivative of any one field in the tilde basis, one has to know
the time derivatives of an infinite number of other fields, each
of which in turn is determined by the time derivatives of an
infinite number of yet other fields. This mess can be disentangled
with the help of the level truncation scheme. At level $N$, what
one can do is consider the fields of level $N$ containing no
$\a^+_{2n}$ excitations. Since these fields can only be sourced by
the time derivatives of fields at a higher level number, in their
truncated equations of motion the $B$ term in eq.\ref{model} is
absent. One can then determine the time derivatives of these
fields and then plug them into the equations of motion of the
fields they source, determine the time derivatives of these other
fields, and proceed this way recursively. Sending the level number
to infinity, the initial value formulation for each field is as in
the previous paragraph. Another approach to this problem is to
note that the $p_+$ dependence in the Siegel gauge equation of
motion occurs in the form $p_+P^+(\p2)$; one can imagine making
another unitary change of basis which diagonalizes $P^+(\p2)$ so
that the equation of motion for a given field will involve the
time derivative of only that field. Such a change of basis can be
achieved by the unitary transformation,
$$V=\exp\left[-ix^-\sum_{n=1}^\infty
(-1)^n(\alpha_{2n}^++\alpha_{-2n}^+)\right]$$ Oddly, operators in
this basis depend explicitly on $x^-$. String field theory is
still translationally invariant, presumably, but now translations
along $x^-$ transform between different modes in the basis. This
seems less palpable to us than the idea that Lorentz
transformations translate between different modes, so we have not
explored this basis seriously.

Solving the dynamical equations of motion in Siegel gauge does not
guarantee that we have fully solved the string field equation. In
addition, there are constraints on the initial conditions at
$x^+(\p2)=0$. To see what these are, multiply the string field
equation by $1$ in the form,
\begin{eqnarray} 0=Q_B\Psi+g\Psi*\Psi &=&
(P^+(\p2))^{-1}[b_0,c(\p2)P^+(\p2)-i\pi_b(\p2)x^+(\p2)']\left(Q_B\Psi+g\Psi*\Psi\right)
\nonumber\\ &=&
(P^+(\p2))^{-1}\left[(c(\p2)P^+(\p2)-i\pi_b(\p2)x^+(\p2)')\left[(L_0+L_0^{gh}-1)\Psi+gb_0\Psi*\Psi\right]\right.
\nonumber\\
&\ &\ \ \ \
+b_0\left.(c(\p2)P^+(\p2)-i\pi_b(\p2)x^+(\p2)')\left[Q_B\Psi+g\Psi*\Psi\right]\right]
\nonumber\end{eqnarray} The first term in this equation is
proportional to the Siegel gauge equations of motion. The second
term has a surprising simplification due to the fact that
$c(\p2)P^+(\p2)-i\pi_b(\p2)x^+(\p2)'$ annihilates interaction
vertex, essentially for the same reason $c(\p2)$ does (see section
5). Therefore, a solution to eq.\ref{Siegel_EOM} is also a
complete solution of the string field equations if and only if,
\begin{equation}b_0\frac{c(\p2)P^+(\p2)-i\pi_b(\p2)x^+(\p2)'}{P^+(\p2)}Q_B\Psi=0\label{gauss}\end{equation}
The thing to notice about this equation is that it is independent
of $p_+$---the $p_+$ dependence cancels because
$[c(\p2)P^+(\p2)-i\pi_b(\p2)x^+(\p2)']^2=0$. Therefore
eq.\ref{gauss} contains no time derivatives and can be interpreted
as a constraint on the initial conditions, like Gauss's law in
electrodynamics. It is remarkable that this constraint is linear
and independent of the string coupling.

Summarizing, the initial value formulation of cubic string field
theory can be described as thus: Fixing Siegel gauge, we can
specify the initial value of the string field at $x^+(\p2)=0$
subject to the constraint eq.\ref{gauss}. Subsequent time
evolution is then simply determined by integrating the Siegel
gauge equations of motion eq.\ref{Siegel_EOM}.

\section{Hamiltonian BRST Quantization}
We now turn to an important application of our formalism:
canonical quantization. Before launching into technicalities,
however, it is important to understand why having a local and
first order formulation of the theory is a crucial element for
defining a consistent and meaningful quantum theory.

For this purpose it is useful to consider how one would
canonically quantize string field theory in the old basis, in
which the Lagrangian depends on an infinite number of time
derivatives. Our discussion reviews that of
ref.\cite{Time-Problem}, and for more details we refer the reader
to that reference. Canonical quantization, of course, proceeds by
translating the theory to the Hamiltonian formalism and replacing
the classical Poisson bracket algebra with an analogous operator
algebra acting on a suitably defined state space. In the case of
string field theory, however, this procedure is complicated by the
fact that the usual Hamiltonian formalism is defined only for
theories whose Lagrangian depends only on coordinates and their
first time derivatives---not an infinite number of their time
derivatives. However there exists a generalization of the
Hamiltonian formalism, due to
Ostrogradski\cite{Ostrogradski,Time-Problem}, which applies to
Lagrangians depending on coordinates and time derivatives up to
any order $N$. For each configuration space coordinate $q$ in this
formalism, there are $2N$ phase space coordinates
$Q_1,Q_2,...,Q_N,P_1,P_2,...,P_N$ representing the $2N$ initial
conditions necessary to specify a solution to the Euler-Lagrange
equations. For string field theory, therefore, every component
field $\phi$ yields $2N$ phase space coordinates with
$N\to\infty$, and each pair $Q_n,P_n$ is associated with a
distinct particle excitation, i.e. each component field generates
an infinite spectrum of particle species. Further, since
$Q_n=(\frac{d}{dt})^{n-1}q$ and $n$ can be arbitrarily large,
field operators at different times commute; $[q(0),q(t)]=0$ for
any finite $t$.

A perhaps even more serious difficulty comes from higher
derivative instabilities. Viewing the Hamiltonian,
$$H=\sum_{n=1}^{N-1}P_n Q_{n+1}+P_N\dot{Q}_N-L(Q_1,...,Q_N,
\dot{Q}_N)$$ one can see that, because $P_n$ and $Q_{n+1}$ are
independent phase space variables, the first term of this equation
can be made arbitrarily negative and the Hamiltonian is unbounded
from below\footnote{By the assumption of nondegeneracy, we can
write $\dot{Q}_N$ as an invertible function of
$Q_1,...,Q_N,P_N$.}. Thus the theory is unstable, possessing
well-known ``runaway'' solutions generic in higher derivative
theories. These instabilities in turn wreak havoc in the quantum
theory. One can attempt to remove the instabilities at the quantum
level by reinterpreting negative energy states with positive norm
as positive energy states with negative norm; but then one either
looses the probabilistic interpretation or one looses unitarity by
removing the negative norm states (which don't decouple) by hand.

These problems are unacceptable in string field theory. It has
been widely felt that somehow the canonical formalism goes awry in
this case, though it has been far from clear how the theory
escapes these sicknesses. Earlier approaches to quantization have
proceeded by defining a configuration space path
integral\cite{Thorn}. Since the Hilbert space structure and
unitarity of the underlying theory are not manifest in this
approach, potential problems with the canonical formalism are
obscured, though presumably nevertheless present. It might be
said, however, that the situation is helped by the fact that the
higher derivatives only appear in the interaction. At the free
level the theory has a standard Hilbert space representation
describing the familiar perturbative string states. The
interaction, since it is included only perturbatively, only adds
small corrections to solutions in an otherwise local and second
order field theory, and in this way problems associated with the
higher derivative nature of the theory do not immediately manifest
themselves.

Our basis, of course, brings new light to the situation. In
midpoint lightcone time, the Lagrangian is local and first order.
The phase space is described in the usual way by the component
fields and their momenta, and the theory should be free of higher
derivative instabilities and related problems at the
non-perturbative interacting level.

So let us construct the canonical quantum theory with the help of
our basis. Since string field theory is a gauge theory there are
many canonical quantization schemes available to us: one may
attempt to quantize without fixing a gauge, as in the Dirac
method; one may attempt to find a unitary gauge, manifestly free
of negative norm states, which would yield a formalism analogous
(though not identical\footnote{Interactions in the lightcone
string field theory are described with a different choice of
interaction vertex, so this formalism cannot be derived by fixing
a gauge in the cubic string field theory alone.}) to lightcone
string field theory; or, one may fix a covariant gauge and
quantize via the Hamiltonian BRST formalism. Here we follow the
last approach, since it is the one seemingly most natural in
covariant string field theory and most closely tied to earlier
quantization schemes which proceeded via the path
integral\cite{Thorn}.

In fixing a covariant gauge, we must in general introduce
Fadeev-Popov ghosts and define a gauge-fixed action with BRST
symmetry. Since cubic string field theory has a complicated
reducible gauge invariance, it is helpful\cite{Thorn} to discuss
ghosts and BRST symmetry in the context of the Batalin-Vilkovisky
(BV) formalism\cite{BV}. For useful reviews of this formalism, see
references \cite{Thorn,Schwartz,BV-Review,Henneaux-Teitelboim}.
The point of departure in the BV formalism is the master action, a
generalization of the usual gauge invariant action which in
addition possesses unphysical ghost fields and antifields and a
BRST symmetry relating them. Gauge fixing the master action yields
the analogy of the Fadeev-Popov gauge fixed action with a residual
BRST symmetry. In the case of cubic string field theory, at least
at tree level, master action takes the same form as the usual
action\cite{Thorn},
\begin{equation}S=-\langle \Psi,Q_B\Psi\rangle
-\fraction{2}{3}g\langle \Psi,\Psi*\Psi\rangle
\label{master}\end{equation} only now, the Grassmann odd string
field $\Psi$ contains as components not only the physical string
field at worldsheet ghost number $1$, but unphysical ghost fields
and antifields at worldsheet ghost numbers $\leq 0$ and $\geq 2$
respectively. The statement that $S$ is BRST invariant at tree
level is expressed by the classical master equation, $$
\{S,S\}=0$$ where $\{,\}$ is a Poisson-like bracket on the
superspace of fields and antifields called the ``antibracket''
(see footnote). At the quantum level BRST invariance is ensured
provided that the master action satisfies the ``quantum'' master
equation,  $$ \{S,S\}=2i\hbar\Delta S$$ where $\Delta$ is a
``symplectic Laplacian'' on the superspace of fields and
antifields\footnote{Explicit formulas for the antibracket and
$\Delta$ are as follows. Given a theory with fields $\varphi^A$
and antifields $\varphi_A^*$ we have $$\{F,G\}= \int dx
F\left(\frac{\overleftarrow{\delta}}{\delta\varphi^A(x)}
\frac{\overrightarrow{\delta}}{\delta\varphi^*_A(x)}-\frac{
\overleftarrow{\delta}}{\delta\varphi^*_A(x)}
\frac{\overrightarrow{\delta}}{\delta\varphi^A(x)}\right)G$$
$$\Delta F = \int dx
\frac{\overrightarrow\delta}{\delta\varphi^*_A(x)}F\frac{\overleftarrow
\delta}{\delta\varphi^A(x)}$$ }.

To proceed we must introduce a basis of states in
$\mathcal{H}_\mathrm{BCFT}$, allowing us to decompose the string
field into an infinite collection of component spacetime fields.
Anticipating the importance of the tilde basis, we define:
\begin{eqnarray}\Phi_i &\equiv& \a_{-l_1}^{\mu_1}...\a_{-l_L}^{\mu_L}\b_{-m_1}
...\b_{-m_M}\c_{-n_1}...\c_{-n_N}|-\rangle'\ \ \ l,m,n\geq 1
\nonumber \\ \Psi_i &\equiv& \c_0\Phi_i\label{basis}\end{eqnarray}
The index $i$ is to be interpreted as a list of indices necessary
to specify the right hand side of eq.\ref{basis}. In this basis an
arbitrary string field can be decomposed as,
$$|A(x)\rangle = A^i(x)\Phi_i+B^i(x)\Psi_i$$
Note that we are working in the position representation, so
$\Phi_i,\Psi_i$ are in the momentum independent component of
$\mathcal{H}_\mathrm{BCFT}$. This basis has definite worldsheet
ghost number, Grassmann parity, and reality properties: \bigskip

\noindent {\it Worldsheet ghost number:}
\begin{equation}gh(\Phi_i)\equiv gh(i)\Phi_i
\ \ \ \ \ \ \ \ \ \
gh(\Psi_i)=(gh(i)+1)\Psi_i\nonumber\end{equation}

\noindent {\it Grassmann parity:}
\begin{equation}\e(\Phi_i)\equiv (-1)^{\e(i)+1}\Phi_i\ \ \ \ \ \
\e(\Psi_i)=(-1)^{\e(i)}\Psi_i\ \ \ \ \ \ \ \ \ \
\e(i)=(gh(i)+1)\mathrm{mod}2\nonumber\\ \label{gh_e}
\end{equation}

\noindent {\it Reality:}
\begin{equation}(\Phi_i,A)\equiv(-1)^{\Im(i)}\Phi_i^+A\ \ \ \
\ \ \ (\Psi_i,A)\equiv(-1)^{\Im(i)+\e(i)}\Psi_i^+A\end{equation}
\bigskip

\noindent Above $(,)$ denotes the momentum independent component
of the BPZ inner product, defined so that $$\langle
A,B\rangle\equiv\int dx(A(x),B(x)) $$ $(,)$ defines an invertible
bilinear form,
\begin{equation}G_{ij}\equiv(\Psi_i,\Phi_j)\label{Gij}\end{equation}
satisfying
\begin{eqnarray} G_{ij}=(-1)^{\e(j)}G_{ji}=G_{ji}(-1)^{\e(i)}&\ &\ \ \ \
\ \ \bar{G}_{ij}=(-1)^{\Im(i)+\Im(j)}G_{ji}\nonumber
\\ gh(i)G_{ij}&=&G_{ij}(2-gh(j))\label{G_id}\end{eqnarray}
As a final bit of notation, consider an operator $\mathcal{O}$ of
Grassmann parity $(-1)^\mathcal{O}$ and independent of ghost zero
modes. We can define an equivalent matrix $\mathcal{O}^i_j$ so
that,
\begin{equation}\mathcal{O}\Phi_i\equiv\mathcal{O}_i^j\Phi_j \ \ \ \ \
\mathcal{O}\Psi_i =
(-1)^{\mathcal{O}}\mathcal{O}_i^j\Psi_j\end{equation} All this
notation will be useful for describing the formalism that follows.

Using the basis $\Phi_i,\Psi_i$ we can decompose the master string
field $\Psi$ in terms of its component fields and antifields.
Write $\Psi$ as a sum of two terms,
$$\Psi=\Psi_-+\Psi_+$$
where $\Psi_-$ contains the physical field and ghosts at
worldsheet ghost numbers $\leq 1$ and $\Psi_+$ contains the
antifields at worldsheet ghost numbers $\geq 2$. We write,
\begin{eqnarray}\Psi_-(x)&=&\Phi_i\phi^i(x)+\Psi_i\psi^i(x)\nonumber\\
\Psi_+(x)&=&\Psi_i
G^{ij}\phi_j^*(x)(-1)^{\e(j)+1}+\Phi_iG^{ij}\psi_j^*(x)(-1)^{\e(j)}\label{fields}\end{eqnarray}
Here we define $\phi^i=\phi_i^*=0$ for $gh(i)\geq 2$ and
$\psi^j=\psi_j^*=0$ for $gh(j)\geq 1$. The string field must be
real and Grassmann odd:
$$\langle \Psi,A\rangle = \Psi^+A\ \ \ \ \ \ \e(\Psi)=-\Psi$$ This
implies,
\begin{eqnarray} \bar{\phi}^i&=&(-1)^{\Im(i)+\e(i)}\phi^i\ \ \ \ \ \ \ \
\e(\phi^i)=(-1)^{\e(i)}\phi^i\nonumber\\
 \bar{\psi}^i&=&(-1)^{\Im(i)+1}\psi^i\ \ \ \ \ \ \ \ \ \
\e(\psi^i)=(-1)^{\e(i)+1}\psi^i\nonumber\\
\bar{\phi}^*_i&=&(-1)^{\Im(i)+\e(i)+1}\phi^*_i\ \ \ \ \ \!
\e(\phi^*_i)=(-1)^{\e(i)+1}\phi^*_i\nonumber\\
\bar{\psi}^*_i&=&(-1)^{\Im(i)}\psi^*_i\ \ \ \ \ \ \ \ \ \ \! \ \ \
\e(\psi^*_i)=(-1)^{\e(i)}\psi^*_i
\end{eqnarray} In particular, note that the fields have opposite
Grassmann parity from their antifields. Finally, we introduce the
notion of (spacetime) ghost number $\mathcal{G}$, defined:
\begin{eqnarray} &\ &\mathcal{G}(\phi^i)=(1-gh(i))\phi^i\ \ \ \ \ \
\mathcal{G}(\psi^i)=-gh(i)\psi^i\nonumber\\
&\ &\mathcal{G}(\phi^*_i)=(gh(i)-2)\phi^*_i\ \ \ \ \ \ \!
\mathcal{G}(\psi^*_i)=(gh(i)-1)\psi^*_i\end{eqnarray} Given that
the string field $\Psi$ and its components satisfy these
properties, following ref.\cite{Thorn} it is straightforward to
show that the master action in eq.\ref{master} has ghost number
zero and satisfies the classical master equation.

Of course, we are ultimately interested in the quantum mechanics,
so it is important to ask whether $S$ as written in
eq.\ref{master} is BRST invariant at the quantum level. However,
transformation to the tilde basis brings a surprise: The action
eq.\ref{master} is automatically BRST invariant at the quantum
level since $\Delta S=0$. This is apparent because $S$ has no term
where any component field is multiplied by its antifield. In the
kinetic term this is obvious since the action has ghost number
zero. In the interaction, this is true because a field associated
with a state $\Phi_i$ is always paired with an antifield
associated with the state $\Psi_jG^{ji}$, or conversely a field
associated with $\Psi_i$ is paired with an antifield associated
with $\Phi_jG^{ji}$. The point is that one of the pair is always
associated with the state $\Psi_i$ which always annihilates the
vertex since it is proportional to $c(\p2)$. Hence, in the tilde
basis fields never couple to their antifields. In the old basis,
however, $\Delta S\neq 0$ and the action apparently must receive
quantum corrections in powers of $\hbar$. These quantum
corrections would manifest themselves in extra Feynmann diagrams
contributing to the loop amplitudes derived from eq.\ref{master}
and in principle ensure gauge invariance of the path integral. The
role of these extra diagrams has been somewhat of a puzzle in
light of well-known arguments\cite{Giddings,Zweibach-Amp} that the
Feynmann diagrams derived from the cubic action eq.\ref{master}
provide a complete and single cover of the moduli space of open
Riemann surfaces, and so in themselves must give the correct open
string theory. Therefore, the absence of quantum corrections in
our formalism provide another justification of the usefulness of
the tilde basis.

We now would like to fix a particular gauge. A convenient
covariant gauge choice is Feynmann-Siegel gauge $b_0\Psi=0$,
analogous to the Landau gauge in Yang-Mills theories. To arrive at
the gauge fixed action, we must add a BRST trivial term to
eq.\ref{master} containing additional fields and antifields and
then eliminate the antifields using a gauge fermion which imposes
delta function gauge fixing conditions in the Siegel gauge. For
details, see ref.\cite{Thorn}. In the end, we find the following
Lagrangian, \begin{eqnarray}L=-\int
dx_\P(\Psi,Q_B\Psi)-\fraction{2}{3}g\int dx_\P (\Psi,\Psi*\Psi) +
2\int dx_\P(\beta,\b_0\Psi)\label{Lag}\end{eqnarray} where $\int
dx_P$ denotes the integral over $x_\perp$ and $x^-$. Since the
interaction is local in lightcone time, the Lagrangian depends
only on the fields and their first time derivatives. Integrating
out the auxiliary field $\beta$ clearly imposes the Siegel gauge
condition. The fields $\Psi$ and $\beta$ can be expanded in terms
of components,
$$\Psi(x)=\Phi_i\phi^i(x)+\Psi_i\psi^i(x)\ \ \ \ \ \ \ \ \beta(x)
=\Psi_i\beta^i(x) $$ The sums over $i$ here now go over all ghost
numbers; for ghost numbers less than or equal to 1 $\phi^i,
\psi^i$ are the same physical string field and ghosts as in
eq.\ref{fields}; for ghost numbers greater than 1, $\phi^i,
\psi^i$ are unphysical antighosts introduced in the gauge fixing
procedure. At any rate, $\phi^i, \psi^i$ satisfy the same ghost
number, Grassmann parity, and reality properties as before. The
action $S_{gf}=\int dx^+ L$ possesses a gauge fixed BRST
symmetry\cite{Thorn},
\begin{eqnarray}s\Psi_- = (Q_B\Psi+\Psi*\Psi)_-\ \ \ \ \ \
s\Psi_+ = -(b_0\beta)_+ \ \ \ \ \ s\beta=0\end{eqnarray} where
$A_-,A_+$ denotes the component of $A$ with worldsheet ghost
numbers less than or equal to 1 and greater than 1, respectively.
Via the Noether procedure we can calculate the conserved spacetime
BRST charge, which takes the simple form,
\begin{eqnarray}\Omega = \int dx_\P (\Psi,
[\c_0(\partial_--iD)-iX]Q_B\Psi)\label{space_BRST}\end{eqnarray}
where we have for shorthand defined, $$ D\equiv\pi P^+(\p2)-p^+\ \
\ \ \ X\equiv -i\pi x^+(\p2)'\pi_b(\p2)$$ In the quantum theory,
of course, $\Omega$ should be nilpotent and define the physical
states through its cohomology.

The Lagrangian eq.\ref{Lag} will be our starting point for
canonical quantization. The first step is to define momenta
$\pi^i,\sigma^i,\chi^i$ canonically conjugate to the component
fields $\phi^i,\psi^i,\beta^i$ respectively. These satisfy the
usual Poisson bracket relations\footnote{From here on, we will
write $x=x_\P$ to avoid unnecessary notational clutter and write
$d=26-1$ for the number of transverse dimensions.},
\begin{eqnarray}\left[\phi^i(x),\pi_j(x')\right]_{\mathrm{PB}}
&=&-(-1)^{\e(i)}\left[\pi_j(x'),\phi^i(x)\right]_{\mathrm{PB}}
=(-1)^{\e(i)}\delta^i_j\delta(x-x')  \nonumber\\
\left[\psi^i(x),\sigma_j(x')\right]_{\mathrm{PB}}&=&
(-1)^{\e(i)}\left[\sigma_j(x'),\psi^i(x)\right]_\mathrm{PB}
=-(-1)^{\e(i)}\delta^i_j\delta(x-x')\nonumber\\
\left[\beta^i(x),\chi_j(x')\right]_\mathrm{PB}
&=&(-1)^{\e(i)}\left[\chi_j(x'),\beta^i(x)\right]_\mathrm{PB}
=-(-1)^{\e(i)}\delta^i_j\delta(x-x')
\end{eqnarray}
Due to our use of lightcone time, it turns out that none of the
momenta are invertible functions of the velocities---i.e. we have
constraints:
\begin{eqnarray}\varphi^1_i&=&G_{jk}[(\partial_-+iD)^j_i\phi^k +
iX^k_i\psi^j]-\pi_i\approx 0\nonumber\\ \varphi^2_i &=&
iG_{jk}X^j_i\phi^k + \sigma_i\approx 0\nonumber\\
\varphi^3_i &=& \chi_i\approx 0\nonumber\\
\varphi^4_i &=& G_{ij}\beta^j \approx 0\nonumber\\ \varphi^5_i &=&
G_{ij}\psi^j \approx 0\end{eqnarray} where we have fixed the gauge
$\beta^i=0$. Solving these constraints we see that the only
independent phase space degrees of freedom are the $\phi^i$s, and
further since the constraints are second class, the $\phi^i$s have
a well-defined Poisson bracket with respect to the induced
symplectic structure on the constraint surface, i.e. the Dirac
bracket:
\begin{equation}[\phi^i(x),\phi^j(x')]_\mathrm{DB}
=\half\left(\frac{1}{\partial_--iD}\right)^i_kG^{kj}\delta(x-x')
\label{class_comm}\end{equation} It is the $\phi^i$s and this
bracket which represents the true canonical structure of the
theory.

The operator appearing in front of the delta function in
eq.\ref{class_comm} is essentially the inverse of $P^+(\p2)$ and
is singular due to the fact that $P^+(\p2)$ has a continuous
spectrum around 0. This is analogous to the situation of a
relativistic scalar field formulated in lightcone frame, where on
the right hand side we find the inverse of $\partial_-$. The
ambiguity in defining the inverse of $\partial_-$ is fixed by
choosing a principal value contour prescription,
$$\frac{1}{\partial_-}\delta(x-x')\equiv\int
\frac{dk}{(2\pi)^d}\mathcal{P}\left(\frac{e^{ik\cdot(x-x')}}{ik_-}\right)
=\half\mathrm{sign}(x^--x^{-'})\delta(x_\perp-x'_\perp)$$ which is
the only choice consistent with the symmetry properties of the
Poisson bracket. In our case, we may define the inverse of
$P^+(\p2)$ by making the Taylor
expansion,$$\left(\frac{1}{\partial_--iD}\right)^i_j\delta(x-x')
=\frac{1}{\partial_-}\sum_{n=0}^\infty
\frac{i^n(D^n)^i_j}{\partial_-^n}\delta(x-x')$$ where again we
take the principal value contour prescription,
$$\frac{1}{\partial_-^n}\delta(x-x')\equiv\int
\frac{dk}{(2\pi)^d}\mathcal{P}\left(\frac{e^{ik\cdot(x-x')}}{(ik_-)^n}\right)
=\frac{1}{2(n-1)!}\mathrm{sign}(x^--x'^{-'})(x^--x^{-'})^{n-1}\delta(x_\perp-x'_\perp)$$
With this definition the right hand side of eq.\ref{class_comm} is
consistent with the symmetry of the Poisson bracket.

We now elevate the classical component fields $\phi^i$ to quantum
operators satisfying the Hermiticity property,
\begin{equation}(\phi^i)^+=(-1)^{\e(i)+\Im(i)}\phi^i\end{equation}
and having ghost number and Grassmann parity,
$$\mathcal{G}(\phi^i)=(1-gh(i))\phi^i\ \ \ \
\e(\phi^i)=(-1)^{\e(i)}\phi^i$$ The correspondence rule says that
the $\phi^i$s satisfy graded commutation relations,
\begin{equation}[\phi^i(x),\phi^j(x')]
=\frac{i}{2}\left(\frac{1}{\partial_--iD}\right)^i_kG^{kj}\delta(x-x')
\label{quant_comm}\end{equation} where, $$[A,B]\equiv
AB-(-1)^{\e(A)\e(B)}BA=-(-1)^{\e(A)\e(B)}[B,A]$$ Dynamical
evolution of the operators $\phi^i$ is determined by Heisenberg's
equations of motion, which at the linear level amount to,
\begin{equation}\left[\partial_+(\partial_--iD)^i_j+(M-\half\partial_\perp^2)^i_j\right]
\phi^j(x,\tilde{x}^+)=0\nonumber\end{equation} where we have
defined $M\equiv \tilde{L}_0|_0+\tilde{L}_0^{gh}-1$. The solution
to these equations can be expanded in Fourier modes,
\begin{equation}\phi^i(x,\tilde{x}^+)=\int \frac{dk}{(2\pi)^{d/2}}e^{ik\cdot x}
\exp\left[i\tilde{x}^+(k_--D)^{-1}(\half k_\perp^2+M)\right]^i_j
a^j(k)\end{equation} A little calculation shows that the mode
operators $a^i(k)$ satisfy the reality properties,
\begin{equation}a^i(k)^+=(-1)^{\e(i)+\Im(i)}a^i(-k)\end{equation}
and commutation relations
\begin{equation}[a^i(k),a^j(k')]
=\half\left(\frac{1}{k_--D}\right)^i_k
G^{kj}\delta(k+k')\label{comm}\end{equation} It is a worthwhile
exercise to prove that the reality properties of both sides of
this equation are consistent. Taking the Hermitian conjugate of
the left hand side gives,
\begin{eqnarray}([a^i(k),a^j(k')])^+&=&(-1)^{\Im(i)+\Im(j)+\e(i)+\e(j)}[a^j(-k'),a^i(-k)]\nonumber\\
&=&(-1)^{\Im(i)+\Im(j)+\e(i)+\e(j)}\half\left(\frac{1}{-k_-'-D}\right)^j_k
G^{ki}\delta(k+k')\nonumber\end{eqnarray} The right hand side
gives,
\begin{eqnarray}\left[\half\left(\frac{1}{k_--D}\right)^i_k
G^{kj}\delta(k+k')\right]^+=\half\left(\frac{1}{k_--\bar{D}}\right)^i_k
(-1)^{\Im(k)}G^{kj}(-1)^{\Im(j)+\e(j)}\delta(k+k')\nonumber\end{eqnarray}
To proceed we must make use of a few properties of $D$. Since $D$
is Hermitian, Grassmann even, and satisfies $(A,DB)=-(DA,B)$, it
has the properties
$$G_{ik}D^k_j=-D^k_iG_{kj}\ \ \ \ \ (-1)^{\Im(i)}\bar{D}^i_j = -
D^i_j(-1)^{\Im(j)} \ \ \ \ \ (-1)^{\e(i)}D^i_j=D^i_j(-1)^{\e(j)}$$
Thus, \begin{eqnarray}\left[\half\left(\frac{1}{k_--D}\right)^i_k
G^{kj}\delta(k+k')\right]^+&=&\half(-1)^{\Im(i)+\Im(j)+\e(j)}\left(\frac{1}{k_-+D}\right)^i_k
G^{kj}\delta(k+k')\nonumber\\
&=&\half(-1)^{\Im(i)+\Im(j)+\e(j)}G^{ik}\left(\frac{1}{k_--D}\right)^j_k
\delta(k+k')\nonumber\\
&=&\half(-1)^{\Im(i)+\Im(j)+\e(j)+\e(i)}\left(\frac{1}{-k_-'-D}\right)^j_kG^{ki}
\delta(k+k')\nonumber\end{eqnarray} So the equation works out
consistently. It is also worth checking that the ghost number
properties of eq.\ref{comm} are consistent. The left hand side has
$\mathcal{G}=0$, whereas the right hand side is,
\begin{eqnarray}\mathcal{G}([a^i(k),a^j(k')])&=&(\mathcal{G}(a^i)+\mathcal{G}(a^j))[a^i(k),a^j(k')]\nonumber\\
&=&(2-gh(i)-gh(j))\half\left(\frac{1}{k_--D}\right)^i_k
G^{kj}\delta(k+k')\nonumber\\
&=&[2-gh(i)-(2-gh(i))]\half\left(\frac{1}{k_--D}\right)^i_kG^{kj}\delta(k+k')=0
\nonumber\end{eqnarray} using eq.\ref{G_id} and the fact that $D$
is ghost number zero. As expected, oscillators only have
nontrivial commutation relations when the sum of their ghost
numbers vanish.

The final ingredient in constructing the quantum theory is
defining a suitable representation of the algebra eq.\ref{comm}.
Note that we can rewrite eq.\ref{comm} in the familiar form,
\begin{equation}[a^i(k),a^j(k')^+]
=\half\left(\frac{1}{k_--D}\right)^i_k
G^{kj}(-1)^{\e(j)+\Im(j)}\delta(k-k')\nonumber\end{equation}
Apparently the $a^i$s satisfy a harmonic oscillator algebra and we
can define the usual Fock space representation by acting creation
operators on a vacuum. However, there is a complication due to the
fact that the creation and annihilation operators are not
independent; the creation operator at momentum $k$ is the same as
the annihilation operator at momentum $-k$. Apparently, it is not
consistent to define a vacuum annihilated by all the $a^i(k)$ for
all $k$, since these operators are non-commuting. Rather, we
define the vacuum to be a state of ghost number zero satisfying,
\begin{equation} a^i(k)|0\rangle=0\ \ \ k_->0, \ \ \
\ \mathcal{G}(|0\rangle)=0\end{equation} so that for example
$a^i(-k)|0\rangle, k_->0$ describes a single string in state $i$
and momentum $k$. In this way, we have a Fock space capable of
describing multiple string states whose minus component of
momentum is strictly positive. The positivity of $k_-$ can be
understood as a consequence of the requirement that the lightcone
energy $k_+$ be positive, since for a state of mass $m$ we have
$k_+=\frac{\frac{1}{2}k_\perp^2+m^2}{k_-}$.

Since the righthand side of eq.\ref{comm} is not a positive
definite bilinear form it is clear that our theory has negative
norm states. To establish unitarity we must consider the role of
the BRST charge eq.\ref{space_BRST}. At the quantum level, the
BRST charge should be nilpotent and define the physical states as
elements of its cohomology. Fortunately, the BRST charge is simply
quadratic in the fields and so there is no ordering ambiguity in
lifting it to the quantum level. In terms of modes, we can write
the BRST operator as,
\begin{equation}\Omega=-i\int dk a^i(k)\Omega_{ij}(k)a^j(-k)\end{equation}
where,
\begin{equation}\Omega_{ij}(k)\equiv
(\Phi_i,[\c_0(k_- +D)+X]Q_B\Phi_j)\end{equation} where $Q_B$ in
this expression is evaluated at momentum $-k$ with $k_+=0$. Let us
now establish explicitly the $\Omega$ is nilpotent. Calculating,
\begin{eqnarray}\Omega^2&=&\half[\Omega,\Omega]\nonumber\\
&=& -\fraction{1}{2}\left[\int dk a^i(k)a^l(-k)\Omega_{il}^2(k)
+\int dk a^j(-k)a^l(k)(-1)^{\e(j)}\Omega_{lj}^2(k)\right]\nonumber
\end{eqnarray} Where,
$$\Omega^2_{il}(k)=\Omega_{ij}(k)\left(\frac{1}{-k_--D}\right)^j_mG^{mk}
\Omega_{kl}(k)$$ Calculating,
\begin{eqnarray}\Omega^2_{il}(k)=(\Phi_i,[\c_0(k_-
+D)+X]Q_B\frac{1}{-k_--D}\Phi_mG^{mk}(\Phi_k,[\c_0(k_-
+D)+X]Q_B\Phi_l))\nonumber\end{eqnarray} Now note that for any
string field $\Psi$ we have the property, \begin{equation}\Phi_m
G^{mk}(-1)^{\e(k)}(\Phi_k,\Psi)=\b_0\Psi\end{equation} Using this
fact we can simplify the right hand side,
\begin{eqnarray}\Omega^2_{il}(k)&=&(\Phi_i,[\c_0(k_+
-D)+X]Q_B\frac{\b_0}{-k_--D}[-\c_0(k_-
+D)+X]Q_B\Phi_l)(-1)^{\e(l)+1}\nonumber\\
&=&(\Phi_i,[\c_0(k_-
+D)+X]Q_B[-\c_0(k_-
+D)-X]\frac{\b_0}{-k_--D}Q_B\Phi_l)(-1)^{\e(l)+1}\nonumber\\
&\ &\ \ +(\Phi_i,[\c_0(k_- +D)+X]Q_B[\b_0,\c_0(k_-
+D)]\frac{1}{-k_--D}Q_B\Phi_l)(-1)^{\e(l)+1}\nonumber\end{eqnarray}
Since $Q_B$ is nilpotent and can be written as
$Q_B=k_+(\c_0(k_-+D)+X)+Q_B|_{k_+=0}$ when $k_+\neq 0$, it follows
that $$[Q_B,\c_0(k_-+D)+X]=0$$ and
therefore,\begin{eqnarray}\Omega^2_{il}(k)&=&(\Phi_i,[\c_0(k_-
+D)+X]^2 Q_B\frac{\b_0}{-k_--D}Q_B\Phi_l)(-1)^{\e(l)+1}\nonumber\\
&\ &\  +(\Phi_i,[\c_0(k_-
+D)+X]Q_B^2\Phi_l)(-1)^{\e(l)+1}\nonumber\end{eqnarray} which
vanishes as a consequence of $Q_B^2=[\c_0(k_- +D)+X]^2=0$.
Therefore, $\Omega$ is nilpotent and defines a cohomology; by the
usual prescription, we identify the physical Hilbert space of the
theory with the state cohomology of the BRST operator at ghost
number zero, $\mathcal{H}_{phys}=H_0(\Omega)$, and likewise
quantum mechanical observables with the operator cohomology at
ghost number zero.

A consistency requirement, of course, is that the physical Hilbert
space should be preserved under dynamical evolution. This would be
guaranteed by the fact that $\Omega$ is a conserved charge, so it
is important to verify this explicitly. The free Hamiltonian takes
the form, \begin{equation}H_0=\int dk
a^i(k)H_{ij}(k)a^j(-k)\end{equation} where,
$$H_{ij}(k)=(\Psi_i,[k_\perp^2+M]\Phi_j)$$
In normal ordered form this can be written, $$H_0 =
2\int_{k_->0}dk a^i(-k)H_{ij}(k)a^j(k) + \half\int_{k_->0}dk
\delta(0)
(M+k_\perp^2)^m_k\left(\frac{1}{k_--D}\right)^k_m(-1)^{\e(m)}$$
The divergent constant term could presumably describe the vacuum
energy of a space-filling D25 brane, though really the value of
this constant is a matter of definition. Note that this is the
only ultraviolet divergence we expect to find in the quantum
string field theory, since nonlocality of the interaction
presumably saves us from further ultraviolet divergences at the
interacting level. This of course is what we expect from string
theory. Anyway, to prove that $\Omega$ is a conserved charge we
must calculate\footnote{From the fact that $\Omega$ is independent
of the open string coupling it is apparent that it should commute
with both the free and interacting Hamiltonians. $\Omega$ commutes
with the interaction simply because the operator $\c_0(k_-+D)+X$
annihilates the vertex (see section 5).},
\begin{eqnarray}[H_0,\Omega]&=&-\frac{i}{2}\left[\int dk a^i(k)a^l(-k)I_{il}(k)
+\int dk a^j(-k)a^l(k)I_{jl}(k)\right]\nonumber\end{eqnarray}
where $$I_{il}(k)=H_{ij}(k)\left(\frac{1}{-k_--D}\right)^j_m
G^{mk}\Omega_{kl}(k)
+\Omega_{ij}(k)\left(\frac{1}{-k_--D}\right)^j_m G^{mk}H_{kl}(k)$$
This quantity can be calculated as follows,
\begin{eqnarray}I_{il}(k)&=&(\Psi_i,[k_\perp^2+M][-k_--D]^{-1}\Phi_m)G^{mk}
(\Phi_k,[\c_0(k_-+D)-X]Q_B\Phi_l)\nonumber\\
&\ &\ \ +(\Phi_i,[\c_0(k_-+D)+X]Q_B[-k_--D]^{-1}\Phi_m)G^{mk}(\Psi_k,[k_\perp^2+M]\Phi_l)\nonumber\\
&=&-(\Phi_i,[k_\perp^2+M][-k_--D]^{-1}[\c_0(k_-+D)+X]Q_B\Phi_l)\nonumber\\
&\ & +
(\Phi_i,[\c_0(k_-+D)+X]Q_B[-k_--D]^{-1}[k_\perp^2+M]\Phi_l)\nonumber\\
&=&-\left(\Phi_i,\left[(k_\perp^2+M)\frac{1}{-k_--D}[\c_0(k_-+D)+X]Q_B\right.\right.\nonumber\\
&\ &\ \ \ \ \ \ \ \ \
+\left.\left.Q_B[\c_0(k_-+D)+X]\frac{1}{-k_--D}(k_\perp^2+M)\right]\Phi_l\right)\nonumber
\end{eqnarray} To proceed note that
$$(k_\perp^2+M)\Phi_i=\b_0Q_B\Phi_i$$ where as before $Q_B$ is understood to
be evaluated at $k_+=0$. Thus we may write,
\begin{eqnarray}I_{il}(k)&=&-\left(\Phi_i,\left[Q_B \b_0\frac{1}{-k_--D}[\c_0(k_-+D)+X]Q_B\right.\right.\nonumber\\
&\ &\ \ \ \ \ \ \ \ \
+\left.\left.Q_B[\c_0(k_-+D)+X]\frac{1}{-k_--D}\b_0Q_B\right]\Phi_l\right)\nonumber\\
&=&-\left(\Phi_i,Q_B[\c_0(k_-+D)+X,\b_0]\frac{1}{-k_--D}Q_B\Phi_l\right)=-(\Phi_i,Q_B^2\Phi_l)=0\nonumber
\end{eqnarray} as expected, establishing that $\Omega$ is a conserved
charge.

\section{Causality} Having found a suitable local time
coordinate for open string field theory, the next natural question
to ask is whether the theory is causal. One might imagine, in
particular, asking whether a localized perturbation of the initial
conditions at $x^+(\p2)=0$ only affects the subsequent time
evolution inside the future lightcone of the perturbation. At
first glance the answer would seem to be ``no,'' since the vertex
in the tilde basis is nonlocal in the transverse directions. A
perturbation at $\tilde{x}$ will therefore affect the value of the
string field at $\tilde{y}$ even when $\tilde{x}$ and $\tilde{y}$
are ``spacelike'' separated: $(\tilde{x}-\tilde{y})^2>0$.
Curiously, however, we cannot conclude from this that the theory
is acausal, since $(\tilde{x}-\tilde{y})^2$ is not actually a
Lorentz invariant quantity. This of course makes sense: The
various components of $\tilde{x}$ have different interpretations
in terms of either the midpoint or center of mass degrees of
freedom, and these components are not related to each other only
by a Lorentz transformation. Thus, in a sense the tilde basis has
made the nature of time transparent at the cost of making
causality obscure; in particular, while there is a well-defined
notion of {\it time ordering} in the tilde basis, there is no
sense in which fields in the tilde basis evaluated at distinct
spacetime points have any definite causal relation to one another.

To discuss causality it seems necessary to use a basis where the
component fields do have definite causal relationship.
Specifically, this entails expanding the string field in a basis
or eigenstates of some position operator $\xi$ satisfying
$[\xi,p]=i$ where $\xi\cdot\xi$ is Lorentz invariant. Of course, a
natural choice of $\xi$ is the string center of mass. However,
since the field equations for this $\xi$ are nonlocal in both
space and time, specifying appropriate initial conditions at $t=0$
in some sense only determines time evolution arbitrarily far in
the future, and the past lightcone of the evolution includes
arbitrarily large sections of the initial value surface.
Therefore, one might claim in some trivial sense that localized
perturbations only effect the time evolution in the future
lightcone, but it is hard to really take this argument seriously
since the initial value problem is not under control; it is not
clear to what extent the field for $t>0$ is either independent or
dynamically determined by the initial conditions at $t=0$.
Apparently, to make a really convincing argument for causality we
must choose a covariant $\xi$ for which the initial value problem
is well-defined. There is only one such $\xi$: the midpoint.

This brings good and bad news. The good news is that the theory in
the midpoint basis is Lorentz invariant and local, so causality is
manifest. The bad news is that the midpoint basis is singular: the
component fields carry infinite energy and the vertex is only
local with a careful choice of regularization. Confronted with
these problems, it seems we can only argue for causality in string
field theory by regulating the midpoint basis, and showing that as
the regulator is removed the theory becomes arbitrarily close to
being manifestly causal. Our task therefore is to find a
consistent regulator. Unfortunately neither or the regulators
$\omega_{2n}=\lambda^n(-1)^n,\ \ \lambda\to 1^-$ or $\omega_{2n}=
(2n)^{-s} (-1)^n,\ \ s\to 0^+$ mentioned in section 2 are
acceptable. The $\lambda$ regulator suffers from an anomaly in
locality vertex\cite{Manes,Morris}, and fields in the $s$
regulator acquire infinite energy at $s=1$, even before the
midpoint $s=0$ is reached.

We will construct a regulator which is closer in spirit to the
approach we have taken to the lightcone midpoint basis. For this
purpose, we must introduce some notation. Consider a complex basis
of spacetime vectors $\lambda^i, \chi^i$ with $i=0,...,D/2-1$
satisfying,\begin{eqnarray}\lambda^0_\mu
v^\mu&=&\frac{1}{\sqrt{2}}(v^0+v^1)\ \ \ \ \ \ \ \ \ \
\chi^0_\mu v^\mu=\frac{1}{\sqrt{2}}(v^0-v^1)\nonumber\\
\lambda^j_\mu v^\mu&=&\frac{1}{\sqrt{2}}(v^{2j}+iv^{2j+1})\ \ \ \
\ \chi^j_\mu v^\mu=\frac{1}{\sqrt{2}}(v^{2j}-iv^{2j+1})\ \ \ \ \
j\geq 1 \end{eqnarray} These satisfy,
\begin{equation}\lambda^i\cdot\lambda^j=\chi^i\cdot\chi^j=0\ \ \ \ \ \ \ \
\lambda^i\cdot\chi^j=\eta^{ij}\ \ \ \ \ \
\eta_{ij}=\mathrm{diag}(-1,1,...1)\end{equation} We will use
$\eta_{ij}$ to raise and lower the $i,j$ indices. Since the
$\lambda^i,\chi^i$ form a basis, they furnish a resolution of the
identity, $$\delta^\mu_\nu=\lambda^\mu_i\chi^i_\nu +
\chi^\mu_i\lambda^i_\nu$$ and satisfy the reality properties,
$$\bar{\lambda}^0=\lambda^0\ \ \ \ \ \bar{\chi}^0=\chi^0\ \ \ \ \
\bar{\lambda}^i=\chi^i\ \ \ \ \ \bar{\chi}^i=\lambda^i\ \ \ \ \
i\geq 1$$ We consider a basis defined by the similarity
transformation,
\begin{equation}U_\sigma=\exp\left[-p\cdot\lambda^i\chi_i\cdot\sum_{n=1}^\infty
\frac{(-\sigma^2)^n}{2n}(\alpha_{2n}-\alpha_{-2n})
-p\cdot\chi^i\lambda_i\cdot\sum_{n=1}^\infty
\frac{(-1)^n}{2n}(\alpha_{2n}-\alpha_{-2n})\right]\
\label{U_reg}\end{equation} where $\sigma<1$. This operator is not
unitary, but for $\sigma\to 1$ it becomes unitary and equal to the
operator eq.\ref{U_wrong} defining the completely midpoint basis.
Note that the sum over the $\lambda$ oscillators is not regulated,
but the sum over the $\chi$ oscillators is. The position operator
defined by eq.\ref{U_reg} will therefore not be covariant, but as
$\sigma\to 1$ it will be. Transforming with $U_\sigma$ in the
usual way we define,
\begin{eqnarray}\check{\alpha}_n&=&U_\sigma\alpha_nU_\sigma^{-1}
=\alpha_n-\cos\fraction{n\pi}{2}p\cdot(\lambda^i\chi_i\sigma^{|n|}+\chi^i\lambda_i)
\ \ \ \ n\neq 0\nonumber\\
\check{p}&=&U_\sigma pU_\sigma^{-1}=p\nonumber\\
\check{x}&=&U_\sigma
xU_\sigma^{-1}=x+i\sum_{n=1}^\infty\frac{(-1)^n}{2n}
(\lambda_i\chi^i\sigma^n+\chi_i\lambda^i)\cdot(\alpha_{2n}-\alpha_{-2n})\nonumber\\
\check{|k\rangle}&=&U_\sigma|k\rangle
\end{eqnarray} This basis satisfies the usual properties,
\begin{eqnarray}[\check{\alpha}^\mu_m,\check{\alpha}^\nu_{-n}]&=&n\delta_{mn}\eta^{\mu\nu}
\ \ \ \ \ \mathrm{bpz}(\check{\alpha}_n)=\check{\alpha}_{-n}
\nonumber\\ \check{\alpha}_n\check{|k\rangle}&=&0\ \ \ n\geq 1  \
\ \ \ \ \ \ [\check{x}^\mu,p_\nu]=i\delta^\mu_\nu\nonumber
\end{eqnarray} but, because $U_\sigma$ is not unitary,
$(\check{\alpha}_n)^+\neq\check{\alpha}_{-n}$. This means in
particular that spacetime fields in this basis are subject to a
rather complicated, nonlocal reality condition. While this is
somewhat bothersome it does not pose a problem as far as causality
is concerned; in the limit $\sigma\to 1$ $U_\sigma$ becomes
unitary and the nonlocality of the reality condition disappears.
Calculating $L_0$ we find,
\begin{equation}L_0=\left(\half+\frac{\sigma^2}{1-\sigma^2}\right)p^2
+p\cdot\sum_{n=1}^\infty
(-1)^n[\lambda^i\chi_i\sigma^{2n}+\chi^i\lambda_i]\cdot(\check{\alpha}_{2n}+\check{\alpha}_{-2n})
+\check{L}_0|_0\end{equation} For any $\sigma<1$ $L_0$ is
well-defined, but as expected there is a pole in the $p^2$ term at
$\sigma=1$.

We now turn to the vertex and establish that it is local in the
$\sigma\to 1$ limit. Transforming to the czech basis we find,
\begin{eqnarray}\langle V_3^m| &=& \kappa\int dk^1 dk^2 dk^3
\delta(k^1+k^2+k^3)\check{\langle +,k^1|}\check{\langle +,k^2|}\check{\langle +,k^3|}\nonumber\\
&\ &\ \ \times\exp\left[-\half V_{00}^{AB}k^A\cdot k^B-
V_{m0}^{AB}\check{a}_{m}^A\cdot k^B -\half V_{mn}^{AB}
\check{a}_m^A\cdot
\check{a}_n^B\right]\nonumber\\
&\ &\ \
\times\exp\left[-k^A\cdot\lambda^i\chi_i\cdot\beta_n(\sigma)(\check{a}_n^{+A}
-\check{a}_{-n}^{+A})-k^A\cdot\chi^i\lambda_i\cdot\beta_n(\check{a}_n^{+A}
-\check{a}_{-n}^{+A})\right]\end{eqnarray} where
$\beta_n(\sigma)=\frac{1}{\sqrt{n}}\cos\fraction{n\pi}{2}\sigma^n$.
Pulling the unregulated factor in $U_\sigma$ through the vertex
first, the quadratic momentum dependence and the coupling of the
$\lambda$ oscillators to the momentum disappears completely; the
calculation is exactly analogous to the one in appendix A. Pulling
the regulated factor through the vertex we find the expression,
\begin{eqnarray}\langle V_3^m| &=& \kappa\int dk^1 dk^2 dk^3
\delta(k^1+k^2+k^3)\check{\langle +,k^1|}\check{\langle +,k^2|}\check{\langle +,k^3|}\nonumber\\
&\ &\ \ \times\exp\left\{-
[V_{m0}^{AB}+(V^{AB}_{mn}+\delta^{AB}\delta_{mn})\beta_n(\sigma)]\check{a}_{m}^A
\cdot\chi^i\lambda_i\cdot k^B -\half V_{mn}^{AB}
\check{a}_m^A\cdot \check{a}_n^B\right\}\nonumber \end{eqnarray}
The momentum dependent factor of course vanishes in the limit
$\sigma\to 1$ as a consequence of the identity eq.\ref{matter_id}
found in appendix A. Therefore, we have identified a nonsingular
set of component spacetime fields evolving according to field
equations which are arbitrarily close to being local.

Morally, one expects that since the interaction is arbitrarily
close to being local, the theory should appear causal in the
$\sigma\to 1$ limit. When we say that it ``appears'' causal, we
really mean the following: One can imagine finding two time
dependent solutions $\Psi$ and $\Psi'$ for $\sigma<1$, where the
initial conditions for $\Psi'$ differ from those of $\Psi$ in some
compact region $R$ on the null plane $x^+(\p2)=0$. The expectation
is that, \begin{equation}\lim_{\sigma\to 1}\Psi(x)=\lim_{\sigma\to
1}\Psi'(x) \ \ \ \ (x-y)^2>0\ \ \mathrm{for}\ \ y\in
R\label{causal?}\end{equation} at least up to a gauge
transformation, since the locality of the theory in this limit
would seem to preclude any information from reaching $x$ from $R$.
However, the limit $\sigma\to 1$ is so singular that it is
admittedly premature to claim that locality of the theory in this
limit truly implies eq.\ref{causal?}. Indeed, it is not clear that
the limit in eq.\ref{causal?} even exists, and probably it can
only be interpreted in a distributional sense. Still the existence
of a local limit provides some evidence, for whatever it's worth,
that the theory is in some sense causal.

Another argument in favor of causality can be mounted with the
help of the Moyal formalism developed by Bars and
collaberators\cite{Bars}. In this framework, the string field in
the matter sector is considered as a local function of the
midpoint and higher mode coordinates:
$\Psi=\Psi[x(\p2),x_{2n},p_{2n}]$ where $x_{2n}$ are the even
Fourier modes of $x(\sigma)$ and $p_{2n}$ are a particular linear
combination of the odd Fourier modes of $p(\sigma)$. Witten's star
product is formally given by computing the Moyal product of the
fields, where $[x_{2m},p_{2n}]_\star = \delta_{mn}$, and the
midpoint coordinates are identified locally. Of course, we know
that the theory in such a language must be singular, but Bars {\it
et al} have developed a convenient and reliable regulator whereby
one essentially truncates the theory to include only a finite
number $N$ of $x_{2n},p_{2n}$ and regulates the linear operator
relating $p_{2n}$ to the odd momentum Fourier modes in a
particular way. The truncated action in this formalism is finite
for any $N<\infty$ and furthermore is both Lorentz invariant and
local in $x(\p2)$. Therefore, at any stage as we take $N\to\infty$
it is clear that we are dealing with a causal theory, and it is
natural to suppose that it continues to be so when the limit is
saturated. This argument has the advantage that causality is
manifest at every stage, but has the disadvantage that the Moyal
regulator deforms the structure of the theory. Furthermore, the
gauge invariance of the truncated theory is not understood. As we
have seen gauge invariance and the initial value formulation in
the exact theory seem to imply that the star algebra is
degenerate, and so probably does not admit a well-defined
representation in terms of a Moyal product.

We should say a few more words about the physical interpretation
of the the causal limit eq.\ref{causal?}. Causality requires that
spacelike separated {\it physical} events should not be
correlated. The string field however is not an observable. There
need be no {\it a priori} constraint on causal propagation of the
string field itself, only on the gauge invariant physical degrees
of freedom it represents. Implicit in this language, however, is
that classical open string field theory possesses local gauge
invariant observables which propagate. Our results imply that open
string field theory has observables which are localized in
(lightcone) time---formally, they correspond to classical BRST
cohomology classes of functionals of the initial
conditions---however it seems unlikely that the theory possesses
observables which are local in all spacetime coordinates. What,
then, is the physical significance of the causal limit we have so
carefully constructed? The answer to this question lies in the
S-matrix, which is the only physical observable we explicitly
understand in string theory. Our proposal is that the S-matrix can
be meaningfully formulated in terms of Green's functions of
interpolating component fields in the czech basis. Due to the
locality of the theory in the $\sigma\to 1$ limit, the component
fields should satisfy local and causal commutation relations in
this limit if we fix a covariant gauge. This limit then will make
manifest locality properties of the Green's functions which lead
directly to the analytic properties of a causal S-matrix.

\section{Conclusion} In this paper we have shown that, if time is
identified with the lightcone component of the string midpoint,
cubic string field theory is local and first order in lightcone
time. Further, since the cubic vertex provides a complete single
cover of the moduli space of open Riemann
surfaces\cite{Giddings,Zweibach-Amp}, the fact that the action is
local in time at tree level is sufficient to guarantee that
locality is not spoiled by quantum corrections. We have taken care
to prove that this result is not a formal artifact; we can expand
the string field in a basis of eigenstates of $x^+(\p2)$, and for
such eigenstates the cubic action is well-defined, gauge
invariant, and local in time. We have also identified a singular
midpoint limit where the action is well defined but arbitrarily
close to being local and manifestly causal.

The existence of a local time coordinate in open string field
theory seems to rely crucially on our choice of interaction
vertex. Certainly, the Witten vertex is the simplest choice but
gives only one way of slicing up the moduli space of open Riemann
surfaces. In a more general decomposition\cite{Tensor}, we would
need not only a cubic vertex, but an an infinite sequence of
higher order vertices at tree level and beyond to recover the
correct moduli space. In general it is not obvious that there
exists a unitary transformation analogous to eq.\ref{U} rendering
all of these vertices local in some time coordinate. This is an
interesting question of principle, though our results indicate
that at any rate observables in open string theory can be
described in terms of at least one set of underlying gauge degrees
of freedom for which the initial value problem is well-defined. In
this sense, nonlocality in time in any formulation of open string
field theory must be ``pure gauge,'' regardless of whether the
nonlocality can be removed by an invertible field redefinition.
The situation with respect to closed string theory, on the other
hand, is far from clear. In closed string field theory there is no
simple choice of overlap type vertex or vertices like Witten's
which covers the moduli space of Riemann surfaces to all
genus\cite{Closed-SFT}. The fact that the Witten vertex is an
overlap was certainly instrumental for us, though it is not
completely clear that overlap type vertices are either a necessary
or sufficient condition for the existence of a local time
coordinate. In the case of closed string theory one might be
encouraged by the fact that both open and closed string
interactions can be formulated (at least perturbatively) in the
lightcone gauge string field theory, where by construction the
interactions are local in lightcone time. However, as mentioned in
the introduction, it is not clear whether locality in this context
derives from the underlying gauge invariant string field theory,
or whether it emerges via the process of
``localization\cite{Time-Problem},'' whereby one restricts
attention to perturbative solutions and derives a perturbatively
equivalent local action from an underlying nonlocal one. Lightcone
string field theory may then be inherently perturbative, and the
fact that lightcone time is well-defined in this context does not
guarantee that it is meaningful nonperturbatively. If this is the
case, then it is possible that time in closed string theory can
only be defined ``holographically'' through open string degrees of
freedom.

It is important to understand that implicit in our formalism is a
proposal for what states should be rightfully considered elements
of the algebra of string fields. The criterion is that the cubic
action for a physical string field should be the same regardless
of whether the field is expressed in a basis of eigenstates of the
string center of mass or a basis of eigenstates of $\tilde{x}$.
The two most prominent examples of string fields we know should be
``physical'' are the perturbative string states and the numerical
solution representing the closed string vacuum. Both of these are
well defined in the tilde basis---we have seen this explicitly for
the perturbative string states, and the closed string vacuum is
trivially well-defined (assuming that it is defined in the center
of mass basis, which it seems to be), since it is a state at zero
momentum. For solutions representing lower dimensional branes the
situation is less clear, but we have no reason to believe that
they are so singular at the midpoint that they don't admit
well-defined representation in the tilde basis. The most prominent
example of a state which should {\it not} be considered
``physical'' in our framework is the identity string field, since
it is crucial for consistency and gauge invariance in the tilde
basis that there are fields which have vanishing star product with
any other string field.

It is interesting to observe that, sofar, it has only seemed
possible to find Hamiltonian formulations of string theory in the
lightcone frame, as for example in light cone string field theory
or Matrix/membrane theory. One might be tempted to claim that our
result fits into this general pattern, and that all of these
formulations indicate that time in string theory can only be
understood in the lightcone. While this may be true, we cannot at
this point conclude that the appearance of lightcone time in all
of these formulations is more than a coincidence. Lightcone time
in membrane theory and lightcone string field theory emerges for a
very different reason: the lightcone allows a tractable solution
of the theory's constraints. In lightcone string theory, lightcone
time is useful even before strings are allowed to interact. We of
course are not trying to solve any constraints, and lightcone time
only appears prominent at the interacting level by considerations
of energetics. However, the common use of lightcone time is
suggestive, and it would be interesting to establish a theoretical
connection. Actually, the connection to lightcone string field
theory may not be so far off, since in fixing lightcone gauge
$x^+(\sigma)=\mathrm{const.}$ one is already using a formalism
where the midpoint (or any $x^+(\sigma)$) plays the role of time.

There are many interesting directions one might explore beyond our
current work. It is worthwhile to continue our preliminary
analysis of the canonical quantum theory, for example to calculate
perturbative amplitudes and establish unitarity. One hope is that
the canonical formalism may provide a novel perspective on the
role of closed strings in open string field theory, though in this
connection it is probably necessary to consider the superstring
due to well-known difficulties with the closed-string tachyon and
BRST invariance in the bosonic theory\cite{Closed}. Of course, our
results for the bosonic theory should carry over directly to the
superstring, in either the RNS\cite{Witten-Super} or the
Berkovitz\cite{Berkovitz} formalism, since their interactions can
be formulated directly in terms of Witten-type vertices. It might
also be useful to have a more explicit understanding of the gauge
invariance of the theory from a Hamiltonian perspective, in
particular to work out the constraint algebra and to explore other
quantization schemes and gauge fixing procedures.

We have spent some time discussing the role of causality in string
field theory, though what we have had to say on this subject is
clearly far from the final word. To begin, it would be nice to
have a more explicit argument relating macroscopic causality of
the S-matrix to the midpoint limit. Since this limit is quite
singular it is not {\it a priori} obvious that such an argument
can be made in a convincing way. However, at a deeper level, it is
interesting to observe that we had to appeal to intuition from
local field theory as a ``crutch'' in order even to discuss
causality in string field theory, forcing us to describe the
theory in a singular and probably inappropriate fashion. Perhaps
it is possible to formulate a more general and less singular
criterion of microscopic causality that could be applied to string
theory. These issues deserve more thought, but we believe that the
ideas presented in this paper, particularly the initial value
formulation, provide a good starting point.

Another crucial avenue to explore is the construction of time
dependent solutions. Some preliminary work in this direction is
underway and will be reported in ref.\cite{me}, though it might be
said that finding and interpreting such solutions presents a
formidable numerical problem. Already at level (2,4) in the tilde
basis there are eight spacetime fields and more than a hundred
nonlinear terms with complicated momentum dependence in the
equations of motion. In addition one must worry about constraints
on the initial conditions and their consistency with time
evolution of the truncated equations of motion. The hope is that
one can find some evidence for the existence of Sen's ``rolling
tachyon'' boundary conformal field theory solution\cite{Sen},
where the tachyon rolls down from the unstable maximum
homogenously towards the closed string vacuum, but does not cross
over in finite time. In the lightcone formalism, however,
homogenous solutions are less natural. Furthermore, for generic
initial conditions, the many fields in the tilde basis will
undergo complicated chaotic motion after they fall off the
unstable maximum, and in general should not be expected to
approach the closed string vacuum at all. Clearly, there are many
challenging problems in this direction which need to be explored.

We would like to thank Wati Taylor for some discussions. The work
of TE was supported by the National Science Foundation under Grant
No. PHY00-98395, and that of DG by National Science Foundation
under Grant No. PHY99-07949.

\begin{appendix}

\section{Interaction Vertex and Gauge Invariance}
The whole point in our change of basis eq.\ref{U} is that the
interaction in this language should contain no lightcone time
derivatives, and hence the dynamical structure of the theory is
defined purely by the kinetic term, which (as in usual field
theory) is first order in $\partial_+$. We now study the
interaction explicitly and show that, with a few reasonable
caveats, this is in fact the case. The interaction in cubic string
field theory is defined by the three string vertex $\langle
V_3|\in\mathcal{H}^*_{\mathrm{BCFT}}
\otimes\mathcal{H}^*_{\mathrm{BCFT}}
\otimes\mathcal{H}^*_{\mathrm{BCFT}}$:
$$-\frac{g}{3}\langle \Psi, \Psi*\Psi \rangle =
-\frac{g}{3}\langle V_3||\Psi\rangle|\Psi\rangle|\Psi\rangle$$
where,\begin{eqnarray}\langle V_3| &=& \langle V_3^m|
V_3^{gh}\nonumber\\
\langle V_3^m| &=& \kappa\int dk^1 dk^2 dk^3
\delta(k^1+k^2+k^3)\langle +,k^1|\langle +,k^2|\langle +,k^3|\nonumber\\
&\ &\ \ \times\exp\left[-\half V_{00}^{AB}k^A\cdot k^B-
V_{m0}^{AB}a_{m}^A\cdot k^B -\half V_{mn}^{AB} a_m^A\cdot
a_n^B\right]\nonumber\\
V_3^{gh} &=& \exp\left[X_{m0}^{AB}c_m^A b_0^B + X_{mn}^{AB}c_m^A
b_n^B\right]
\end{eqnarray} Here, $A=1,2,3$ denotes which copy of
$\mathcal{H}^*_\mathrm{BCFT}$ the oscillator acts on and $m,n =
1,2,...\infty$ denotes the mode number of the oscillator, repeated
indices summed. Here we use normalized oscillators in the matter
sector $a^\mu_n = \frac{1}{\sqrt{|n|}}\alpha^\mu_n, n\neq 0$ and
$\alpha'=\half$. Explicit expressions for the constants $\kappa,
V$, and $X$ (the latter two are called ``Neumann coefficients'')
were calculated in ref.\cite{Gross-Jevicki} and can be found for
convenient reference for example in ref.\cite{Taylor-Review}. Note
that the matter part of the vertex contains a factor $$\exp[\half
V_{00}^{AB}k_0^A k_0^B - V_{m0}^{AB}a^{0A}_m k_0^B]$$ and since in
position space $p_0=-i\frac{\partial}{\partial t}$ the vertex
contains time derivatives to an arbitrarily high order which seem
to render the initial value problem for string field theory
ill-defined.

Let us now transform the vertex to the tilde basis. We have in the
matter sector,
\begin{eqnarray}\langle V_3^m| &=& \kappa\int dk^1 dk^2 dk^3
\delta(k^1+k^2+k^3)\langle +,k^1|'\langle +,k^2|'\langle +,k^3|'\nonumber\\
&\ &\ \ \times\exp\left[-\half V_{00}^{AB}k^A\cdot k^B-
V_{m0}^{AB}\tilde{a}_{m}^A\cdot k^B -\half V_{mn}^{AB}
\tilde{a}_m^A\cdot
\tilde{a}_n^B\right]\nonumber\\
&\ &\ \ \times\exp\left[-k_+^A\beta_n(\tilde{a}_n^{+A}
-\tilde{a}_{-n}^{+A})\right]U_{gh}^1U_{gh}^2U_{gh}^3\nonumber
\end{eqnarray} where $\beta_n =
\frac{1}{\sqrt{n}}\cos\frac{n\pi}{2}$. Pulling the factor on the
third line through the factor on the second line, we find:
\begin{eqnarray}\langle V_3^m| &=& \kappa\int dk^1 dk^2 dk^3
\delta(k^1+k^2+k^3)\langle +,k^1|'\langle +,k^2|'\langle +,k^3|'\nonumber\\
&\ &\ \ \times\exp\left[(V_{00}^{AB}+V_{m0}^{AB}\beta_m)k_+^Ak_-^B
-(V_{m0}^{AB}
+(V_{mn}^{AB}+\delta^{AB}\delta_{mn})\beta_n)\tilde{a}_{m}^{+A}
k_+^B\right]\nonumber\\
&\ &\ \ \exp\left[-\half V_{00}^{AB}k_\perp^A\cdot k_\perp^B-
V_{m0}^{AB}\tilde{a}_{m}^A\cdot k_\M^B -\half V_{mn}^{AB}
\tilde{a}_m^A\cdot
\tilde{a}_n^B\right]U_{gh}^1U_{gh}^2U_{gh}^3\nonumber
\end{eqnarray} The factor on the second line contains all of the
lightcone time derivatives in the vertex. Presumably, if the
following identities are satisfied: \begin{eqnarray} 0 &=&
V_{00}^{AB}+V_{m0}^{AB}\beta_m\nonumber\\ 0 &=& V_{m0}^{AB}
+(V_{mn}^{AB}+\delta^{AB}\delta_{mn})\beta_n\label{matter_id}\end{eqnarray}
where $B$ is contracted with a conserved quantity, the vertex
contains no derivatives with respect to lightcone time. Now
consider the ghost component of the vertex.
\begin{eqnarray}\langle V_3| = \langle
V_3^m|(U^1_{gh}U^2_{gh}U^3_{gh})^{-1} \exp\left[X_{m0}^{AB}\c_m^A
b_0^B + X_{mn}^{AB}\c_m^A
\b_n^B\right]\exp\left[-b_0^A\gamma_n(\c_n^A
+\c_{-n}^A)\right]\nonumber\end{eqnarray} where
$\gamma_n=\cos\frac{n\pi}{2}$. Pulling the third factor though the
second factor,
\begin{eqnarray}\langle V_3| =
\langle V_3^m|(U^1_{gh}U^2_{gh}U^3_{gh})^{-1}
\exp\left[(X_{m0}^{AB}+(X_{mn}^{AB}+\delta^{AB}\delta_{mn})\gamma_n)\c_m^A
b_0^B + X_{mn}^{AB}\c_m^A \b_n^B\right]\nonumber\end{eqnarray}
Suppose that
\begin{equation}0=X_{m0}^{AB}+(X_{mn}^{AB}+
\delta^{AB}\delta_{mn})\gamma_n.\label{ghost_id}\end{equation} In
this case, the dependence on $b_0$ disappears from the vertex. In
particular, we have the property anticipated earlier,
$$\langle V_3|\c_0^A=0$$ since $\c_0$ passes unimpeded through the
exponential and acts directly on the $\langle +,k|'$ vacuum which
it annihilates. Altogether, the three string vertex in the tilde
basis takes the form,
\begin{eqnarray}\langle V_3| &=& \kappa\int dk^1 dk^2 dk^3
\delta(k^1+k^2+k^3)\langle +,k^1|'\langle +,k^2|'\langle
+,k^3|'\nonumber\\ &\ &\exp\left[-\half V_{00}^{AB}k_\perp^A\cdot
k_\perp^B- V_{m0}^{AB}\tilde{a}_{m}^A\cdot k_\M^B -\half
V_{mn}^{AB} \tilde{a}_m^A\cdot \tilde{a}_n^B + X_{mn}^{AB}\c_m^A
\b_n^B\right]\label{our_vertex} \end{eqnarray} assuming the
identities eq.\ref{matter_id} and eq.\ref{ghost_id}. Observe that
the vertex is still nonlocal in the transverse and $x^-$
directions (in particular we have a term $k_-^A\tilde{a}^{-B}$).
Because the interaction contains an infinite number of derivatives
with respect to $x^-$ the theory is still nonlocal in ordinary
time $t=\frac{1}{\sqrt{2}}(x^++x^-)$. Only in $\tilde{x}^+$ have
we achieved a well-defined initial value formulation.

Apparently, our entire approach rests on the validity of the
identities eq.\ref{matter_id} and eq.\ref{ghost_id}. These
identities are nontrivial and at best can be expected to hold only
over an appropriately defined domain. It is therefore worthwhile
to study these relations carefully and demonstrate whether or not
they should be taken seriously. Before we launch into a more
explicit and technical analysis, however, it is worth noting that
the relations eq.\ref{matter_id} and eq.\ref{ghost_id} can both be
derived straightforwardly as a consequence of the overlap
conditions: $$\langle V_3|(X^A(\p2)-X^{A+1}(\p2))=0\ \ \ \ \
\langle V_3|c^A(\p2)=0.$$ Since the vertex was explicitly
constructed in ref.\cite{Gross-Jevicki} so that half string
overlap conditions (of which this is a special case) would be
satisfied, barring unanticipated subtleties we expect
eq.\ref{matter_id} and eq.\ref{ghost_id} to hold.

We now turn to a more detailed study of eq.\ref{matter_id} and
eq.\ref{ghost_id}. For this purpose, it is useful to rewrite the
identities in a slightly different form. Define
\begin{eqnarray} M^{AB}_{mn} &\equiv& (-1)^mV^{AB}_{mn} \nonumber\\
\tilde{M}^{AB}_{mn} &\equiv &
-(-1)^m\frac{1}{\sqrt{m}}X^{AB}_{mn}\sqrt{n} \nonumber\\
m^{AB}_n &\equiv& (-1)^nV^{AB}_{n0}\nonumber\\
\tilde{m}^{AB}_n &\equiv& -(-1)^n\frac{1}{\sqrt{n}}X^{AB}_{n0}
\label{Ms}\end{eqnarray} These satisfy the relations,
$$M^{AB}=M^{A+1,B+1}\ \ \ \ M^{AB}_{mn}=(-1)^m M^{BA}_{mn}\ \ \ \
M_{mn}+M^{12}_{mn}+M^{21}_{mn}=\delta_{mn}$$ and
$$m^{AB}=m^{A+1,B+1}\ \ \ \ m^{AB}_n=(-1)^n m^{BA}_n\ \ \ \
m_n+m^{12}_n+m^{21}_n=0$$ with corresponding relations in the
ghost sector\footnote{Note that these relations fix the ambiguity
in the zero mode part of the matter Neumann coefficients due to
momentum conservation. In particular,
$V^{AB}_{00}=\half\ln\frac{27}{16}\delta^{AB}$.}. Here we write
$M\equiv M^{11}$ and $m\equiv m^{11}$ and the sum $A+1$ is taken
mod 3. Reexpressing eq.\ref{matter_id} and eq.\ref{ghost_id} in
terms of $M,m,\tilde{M},\tilde{m}$ and using these relations, we
find five independent identities:
\begin{eqnarray}0 &=& 3m_{2m} +
(3M_{2m,2n}+\delta_{2m,2n})\beta_{2n} \label{matter_e}\\ 0 &=&
m^{12}_{2m-1} + M^{12}_{2m-1,2n}\beta_{2n}\label{matter_o}\\ 0 &=&
\ln\frac{27}{16} + 3m_{2n}\beta_{2n}\label{matter_0}\\ 0 &=&
\tilde{m}_{2m} + (\tilde{M}_{2m,2n} - \delta_{2m,2n})\beta_{2n}
\label{ghost_e}\\ 0 &=& \tilde{m}^{12}_{2m-1} +
\tilde{M}^{12}_{2m-1,2n}\beta_{2n} \label{ghost_o}
\end{eqnarray} We will call these equations the ``midpoint identities.''
Some of these relations have appeared elsewhere in the literature.
For instance, eq.\ref{ghost_e} plays an important role in vacuum
string field theory, where it was used to prove\cite{Okuyama2}
that the pure ghost kinetic operator first proposed by Hata and
Kuwano\cite{Hata} based on their Siegel gauge solution is
equivalent to the midpoint value of the $c$ ghost.

We now present detailed analytic evidence in support of the
midpoint identities. Our approach uses the spectrum of the Neumann
coefficients as described in ref.\cite{Spectroscopy}. The spectrum
of the Neumann coefficients is defined by an orthonormal basis of
eigenfunctions $v_n(\k)$,
\begin{equation}\sum_{n=1}^\infty v_n(\k)v_n(\k') = \delta(\k-\k')\ \
\ \ \int_{-\infty}^{\infty} v_m(\k)v_n(\k) = \delta_{mn}
\end{equation} Up to normalization, the $v_n(\k)$s are a complete set
of orthogonal polynomials of degree $m-1$ and satisfy
$v_n(-\k)=(-1)^{n+1}v_n(\k)$\footnote{For a convenient reference
on the various properties of these functions, see the appendix of
ref.\cite{Erler2}.}. They can be defined implicitly via the
generating function,
\begin{equation}\sum_{n=1}^\infty\frac{z^n}{\sqrt{n}}v_n(\k) = \frac{1}{\k N(\k)}
(1-e^{-\k \tan^{-1}z})\label{v_gen}\end{equation} where,
$$N(\k)=\sqrt{\frac{1}{\k}\sinh\fraction{\pi\k}{2}}$$ The
$v_n(\k)$s satisfy the crucial
properties\cite{Spectroscopy,Erler1},
\begin{eqnarray}M_{mn}v_n(\k)&=&-\frac{1}{1+2\cosh\frac{\pi\k}{2}}v_m(\k)\nonumber\\
M^{12}_{mn}v_n(\k)&=&-\frac{1+\cosh\frac{\pi\k}{2}+\sinh\frac{\pi\k}{2}}{1
+2\cosh\frac{\pi\k}{2}}v_m(\k)\nonumber\\
M^{21}_{mn}v_n(\k)&=&-\frac{1+\cosh\frac{\pi\k}{2}-\sinh\frac{\pi\k}{2}}{1
+2\cosh\frac{\pi\k}{2}}v_m(\k)\nonumber\\
\tilde{M}_{mn}v_n(\k)&=&\frac{1}{2\cosh\frac{\pi\k}{2}-1}v_m(\k)\nonumber\\
\tilde{M}^{12}_{mn}v_n(\k)&=&\frac{\cosh\frac{\pi\k}{2}+\sinh\frac{\pi\k}{2}
-1}{2\cosh\frac{\pi\k}{2}-1}v_m(\k)\nonumber\\
\tilde{M}^{12}_{mn}v_n(\k)&=&\frac{\cosh\frac{\pi\k}{2}-\sinh\frac{\pi\k}{2}
-1}{2\cosh\frac{\pi\k}{2}-1}v_m(\k)
\end{eqnarray} and are thus eigenvectors of the $M$s.

Using $v_n(\k)$ to transform the midpoint identities
eq.\ref{matter_e}-\ref{ghost_o} into the diagonal basis, we run
into an immediate problem: $\beta_{2m}$ is undefined in this
basis. Explicit calculation shows:
$$\beta(\k)\equiv\sum_{n=1}^\infty
\frac{(-1)^n}{\sqrt{2n}}v_{2n}(\k)=\frac{1}{\k
N(\k)}(1-\cos\infty)$$ Divergences of this sort are by now well
understood in the $\k$ basis\cite{Erler2} and are related to the
delicate role of the midpoint in the vertex, for example with
anomalies in the associativity of the star product. Since our
formalism makes use of the midpoint structure of the vertex in a
crucial way, the appearance of such a divergence is not
surprising. However, clearly we need to keep track of this
singularity and show that it does not endanger the validity of
eq.\ref{matter_e}-\ref{ghost_o} and the proposed form of the
vertex. We can regulate $\beta$ as follows: \begin{equation}
\beta_\omega(\k)\equiv\sum_{n=1}^\infty
\frac{(i\tanh\omega)^{2n}}{\sqrt{2n}}v_{2n}(\k)=\frac{1}{\k
N(\k)}(1-\cos\omega\k)
\end{equation} In the limit $\omega\to\infty$ we should recover
$\beta_{2n}$, but $\beta_\omega(\k)$ does not converge in the
sense of functions. It does, however, converge in the sense of
distributions,
\begin{equation}\lim_{\omega\to\infty}\beta_\omega(\k) =
\frac{1}{N(\k)}\mathcal{P}\left(\frac{1}{\k}\right)\end{equation}
where $\mathcal{P}$ denotes the principal value. Transforming our
identities to the $\k$ basis (with the exception of
eq.\ref{matter_0}) we find,
\begin{eqnarray}0 &=&3m(\k)+2\frac{\cosh\frac{\pi\k}{2}-1}{1
+2\cosh\frac{\pi\k}{2}}\frac{1}{N(\k)}\mathcal{P}\left(
\frac{1}{\k}\right) \label{matter_ek}\\ 0 &=&m^{12}(\k)+
\frac{\sinh\frac{\pi\k}{2}}{1
+2\cosh\frac{\pi\k}{2}}\frac{1}{N(\k)}\mathcal{P}\left(
\frac{1}{\k}\right) \label{matter_ok}\\
0&=&\tilde{m}(\k)+2\frac{1-\cosh\frac{\pi\k}{2}}{2\cosh\frac{\pi\k}{2}
-1}\frac{1}{N(\k)}\mathcal{P}\left( \frac{1}{\k}\right) \label{ghost_ek}\\
0&=&\tilde{m}^{12}(\k)+\frac{\sinh\frac{\pi\k}{2}}{2\cosh\frac{\pi\k}{2}
-1}\frac{1}{N(\k)}\mathcal{P}\left( \frac{1}{\k}\right)
\label{ghost_ok}
\end{eqnarray} where, \begin{equation}m(\k)=\sum_{n=1}^{\infty}
v_{2n}(\k)m_{2n}\label{m_k}\end{equation} and likewise for
$m^{12}(\k), \tilde{m}(\k)$, and $\tilde{m}^{12}(\k)$. Notice that
each midpoint identity in the $\kappa$ basis appears with a
principal value distribution multiplied by a function which
vanishes at $\k=0$. Since there is no pole there would seem to be
no need to specify a contour prescription. The first terms in
these equations, in fact, do not involve the principal value. For
example, upon calculating $m^{12}(\k)$ one finds that
eq.\ref{matter_ok} reads,
\begin{equation}0 = \frac{1}{N(\k)(1
+2\cosh\frac{\pi\k}{2})}\left[\sinh\fraction{\pi\k}{2}\mathcal{P}\left(\frac{1}{\k}\right)
-\frac{\sinh\frac{\pi\k}{2}}{\k}\right]\label{vanish?}\end{equation}
Since there is no pole at $\k=0$ it is very tempting to claim that
this equation is exactly true. However, we should more precisely
think of this equation as holding in the sense of
distributions---when the right hand side is integrated against a
smooth test function, the resulting integral should always vanish.
If we were to choose a test function which is not smooth at
$\k=0$, the validity of eq.\ref{vanish?} is much more dubious.
Suppose we multiply eq.\ref{vanish?} by a delta function.
$$\frac{1}{3\sqrt{\pi}}\left[\delta(\k)\sinh\fraction{\pi\k}{2}\mathcal{P}
\left(\frac{1}{\k}\right)-\frac{\pi}{2}\delta(\k)\right]$$ This
expression involves a product of distributions and is hence
ill-defined. In particular, if we happen to multiply the
$\sinh\frac{\pi\k}{2}$ with the principal value distribution
first, then $\sinh\frac{\pi\k}{2}\mathcal{P}(1/\k)
=\sinh\frac{\pi\k}{2}/\k$ and the expression vanishes as expected.
If however we are unwise enough to multiply $\sinh\frac{\pi\k}{2}$
with the delta function first, the first term vanishes and cannot
cancel the second term; we get a nonzero answer,
$-\frac{\sqrt{\pi}}{6}\delta(\k)$. In the mode basis, the delta
function corresponds to the vector, \begin{equation}\delta_n =
\frac{1}{\sqrt{\pi}}\frac{\sin\frac{n\pi}{2}}{\sqrt{n}},\ \ \ \ \
\ \ \ \ \ \sum_{n=1}^\infty \delta_{n}v_{n}(\k) =
\delta(\k)\label{delta}\end{equation} Multiplying eq.\ref{vanish?}
by a delta function corresponds to evaluating,
\begin{equation}\delta_{2m-1}m^{12}_{2m-1}+\delta_{2m-1}
M^{12}_{2m-1,2n}\beta_{2n}=?\label{amb_id}\end{equation} If our
identities were exactly true, this expression would vanish. In
reality, however, it is ill-defined because the double sum is
ambiguous---the answer depends on which order the summation is
carried out. These problems can be avoided by a simple restriction
of domains. In particular, the midpoint identities should be
interpreted as relations between distributions in the topological
dual of an appropriately defined Hilbert space. For equations
\ref{matter_o} and \ref{ghost_o}, we can take the Hilbert space to
be simply that of square summable sequences $\ell^2$, since
$\delta_n$ does not have finite norm under this inner product. For
equations \ref{matter_e} and \ref{ghost_e}, we would need a
slightly more singular distribution to cause ambiguities, for
example the derivative of a delta function. In the mode number
basis, the derivative of a delta function corresponds to,
$$\delta'_n =
-\frac{1}{\sqrt{\pi}}\frac{\cos\frac{n\pi}{2}}{\sqrt{n}}
\sum_{m=1}^{n/2}\frac{1}{2m-1}\ \ \ \ \ \ \ \ \sum_{n=1}^\infty
\delta'_{n}v_{n}(\k) = \delta'(\k)$$ Apparently it is sufficient
to take the Hilbert space of equations \ref{matter_e} and
\ref{ghost_e} to be $\ell^2$ as well, though we could probably
slightly weaken this requirement since $\delta'_n$ has a norm even
more divergent than $\delta_n$.

\begin{figure}[top]
\begin{center}\resizebox{2.9in}{1.9in}{\includegraphics{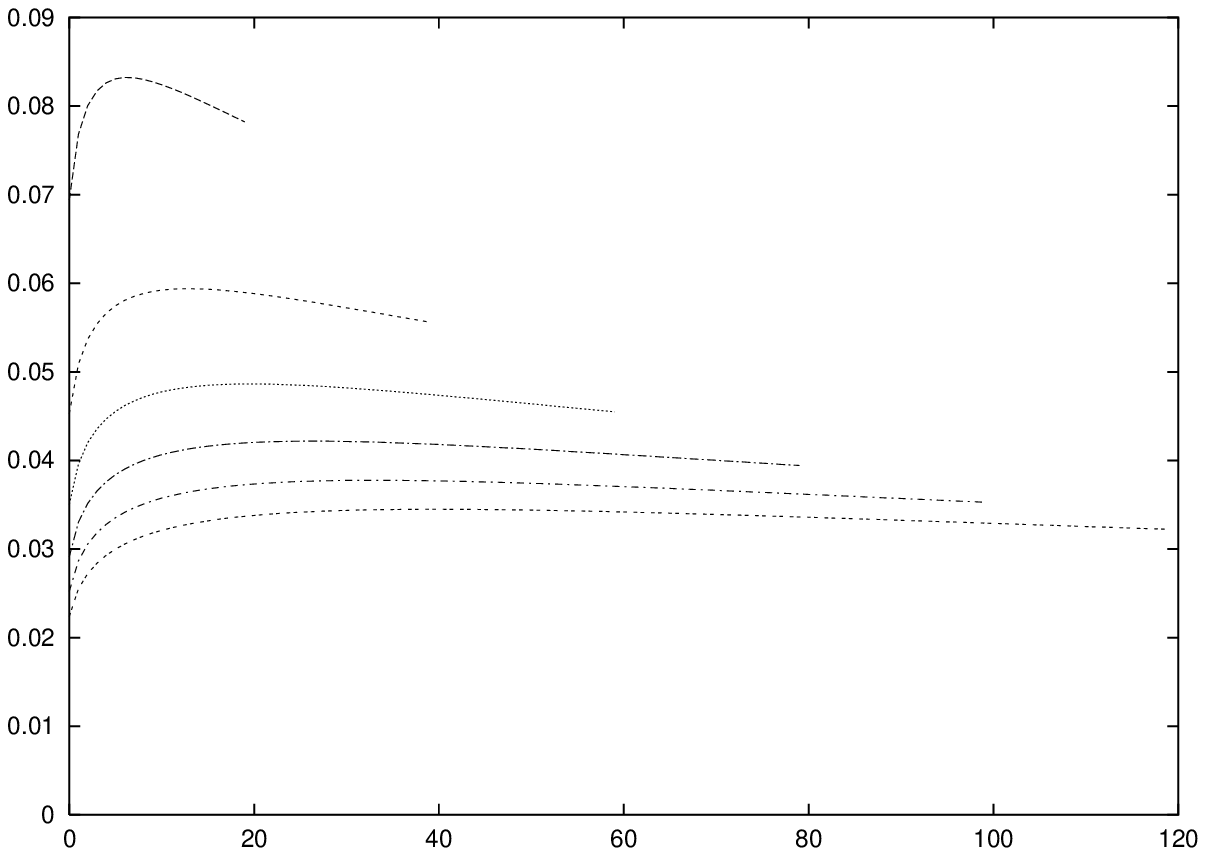}}
\resizebox{2.9in}{1.9in}{\includegraphics{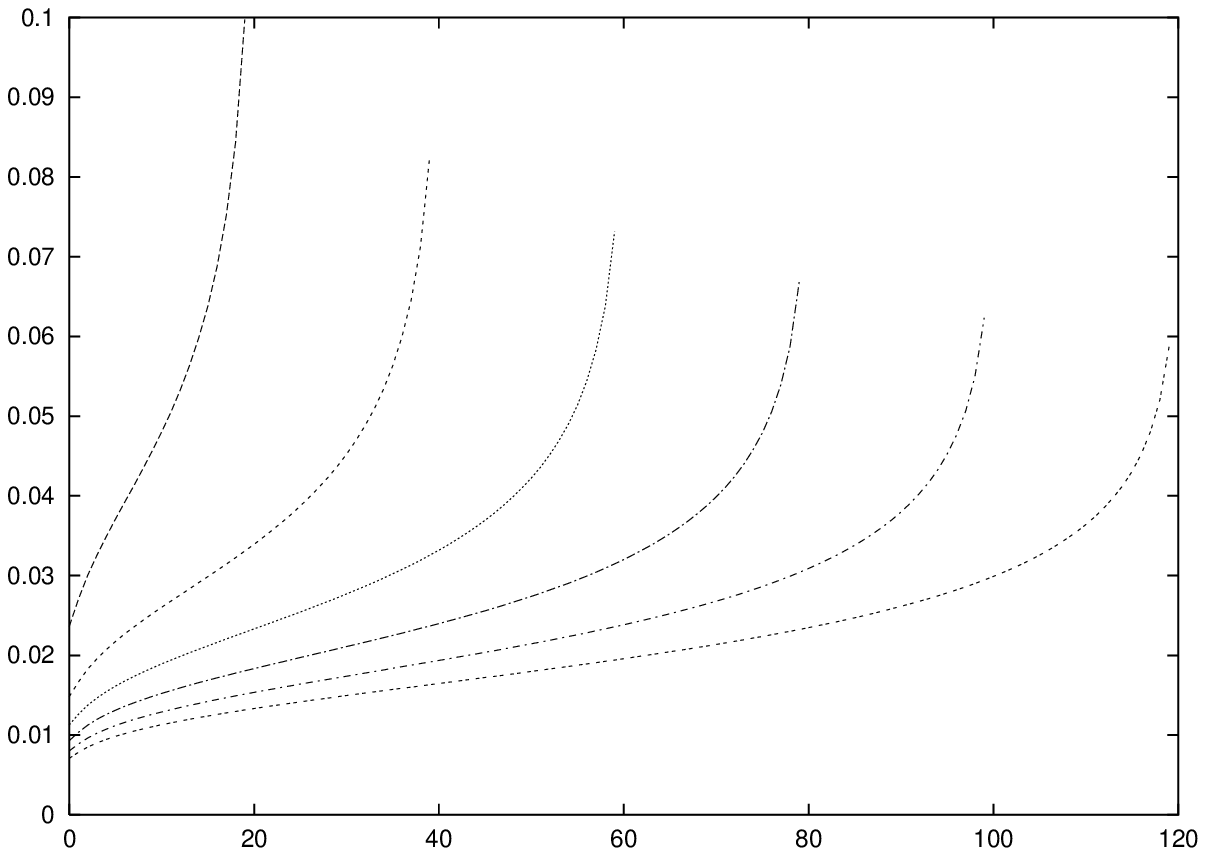}}
\resizebox{2.9in}{1.9in}{\includegraphics{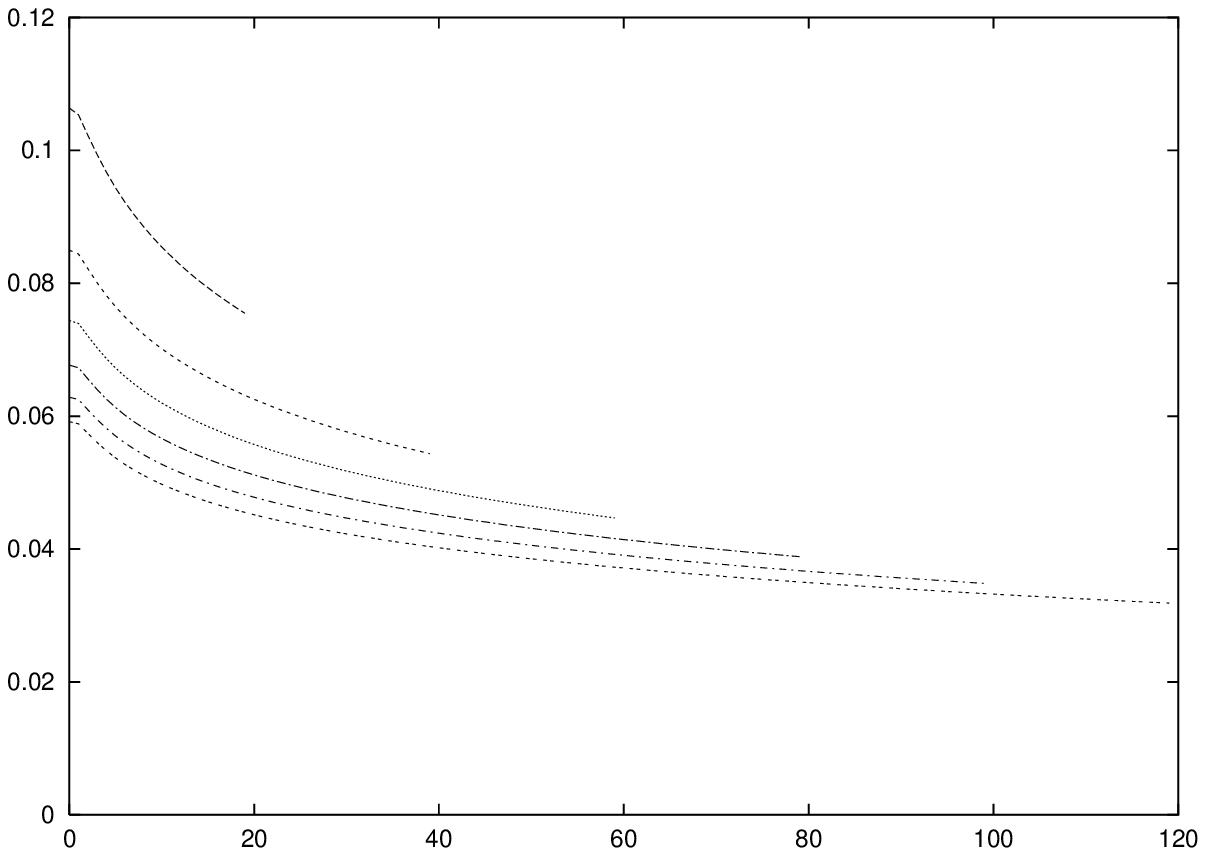}}
\resizebox{2.9in}{1.9in}{\includegraphics{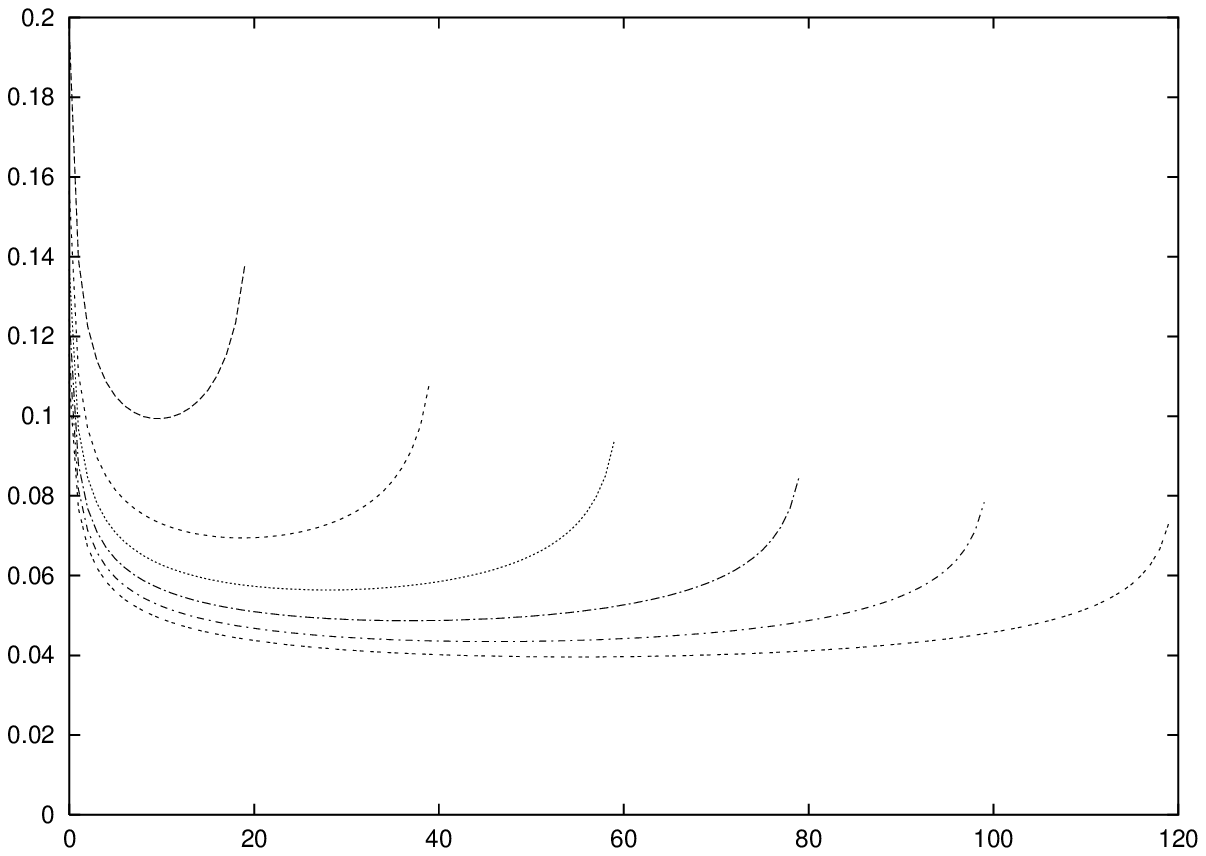}}
\end{center}
Figure 2: The components of $(-1)^n v_n^L$ graphed as a function
of $n$ for $L=20,40,...,120$. Clockwise we have $(-1)^n v_n^L$ for
equations \ref{matter_e}, \ref{matter_o}, \ref{ghost_o},
\ref{ghost_e}
\end{figure}

It is important to realize that our restriction on the Hilbert
spaces translates directly into a restriction on the operators and
string fields for which the vertex can be expected to take the
proposed form eq.\ref{our_vertex}. Let us consider an example of
an operator whose action on the vertex in the old basis does not
agree with with its action on the transformed vertex: the $+$
component of the half string momentum:
\begin{eqnarray}P^{1/2}_+ &=& \int_0^{\pi/2} d\sigma P_+(\sigma) = \half p_+ + \frac{1}{\sqrt{\pi}}
\sum_{n=1}^\infty
\frac{(-1)^n}{2n-1}(\alpha^-_{2n-1}+\alpha^-_{-2n+1})\nonumber\\
&=&\half p_+ -\delta_n(a^-_n+a^-_{-n})=\tilde{P}^{1/2}_+
\end{eqnarray} If we don't assume the identity eq.\ref{matter_o},
the vertex takes the form, \begin{eqnarray}\langle V_3|&=&\langle
V_3^{(us)}|A\nonumber\\
A&=&\exp\left[(m^{12}_{2m-1}+M^{12}_{2m-1,2n}\beta_{2n})(
\tilde{a}^{+1}_{2m-1}p^2_+ -\tilde{a}^{+1}_{2m-1}p^3_+
+\mathrm{cyclic})\right]\nonumber\end{eqnarray} where $\langle
V_3^{(us)}|$ is the vertex of eq.\ref{our_vertex} and ``cyclic''
denotes the sum of $\tilde{a}^{+1}_{2m-1}p^2_+
-\tilde{a}^{+1}_{2m-1}p^3_+$ with the state space indices
cyclically permuted. If the midpoint identities and postulated
vertex are exactly correct, then $A=1$ and $P_+^{1/2}$ should
commute with $A$. However, calculation shows $$[A,P_+^{1/2}]
=(\delta_{2m-1}m^{12}_{2m-1}+\delta_{2m-1}
M^{12}_{2m-1,2n}\beta_{2n})(p_+^2-p_+^3)A$$ We find the same
ambiguous double sum as in eq.\ref{amb_id}. If the sum over $n$ is
performed first the commutator vanishes as expected, but if the
sum over $m$ is performed first, the commutator does not vanish.
This is an indication that the action of $P^{1/2}_+$ on the vertex
does not commute with the transformation into the tilde basis.
Actually, the difficulty with $P^{1/2}_+$ is not surprising since
in an early paper\cite{Horowitz} Horowitz and Strominger showed
that the operator $P^{1/2}$ could be used to generate translations
of the string center of mass (and hence $\tilde{x}$) of a string
field using the star product. In particular, $P^{1/2}_+$ could be
used to translate $\tilde{x}^+$. This, however, is inconsistent
with the conjectured locality of the vertex in lightcone time:
Evaluating the star product of two fields at a particular
$\tilde{x}^+$ will always generate another field at the {\it same}
$\tilde{x}^+$. In the ghost sector, similar difficulties are
encountered with $b(\p2)$. We are fortunate that these operators
do not seem to play a crucial role in the theory---they do not
appear in the BRST operator, BPZ inner product, or the Virasoros
expressed in the tilde basis. We therefore expect that the
proposed vertex eq.\ref{our_vertex} should be valid for a
sufficiently general and physically interesting class of string
fields.

Modulo issues of domains, we still need to prove the that the
$m(\k)$s cancel the principal value terms in
eq.\ref{matter_ek}-\ref{ghost_ok} as expected. This is easily done
along the lines of Okuyama in his proof\cite{Okuyama2} that the
kinetic operator $\mathcal{Q}$ of vacuum string field theory is
$c(\p2)$. For completeness we have included a sample computation
in appendix B, along with a separate derivation of
eq.\ref{matter_0}.

\begin{figure}[top]
\begin{center}a)\resizebox{2.7in}{1.7in}{\includegraphics{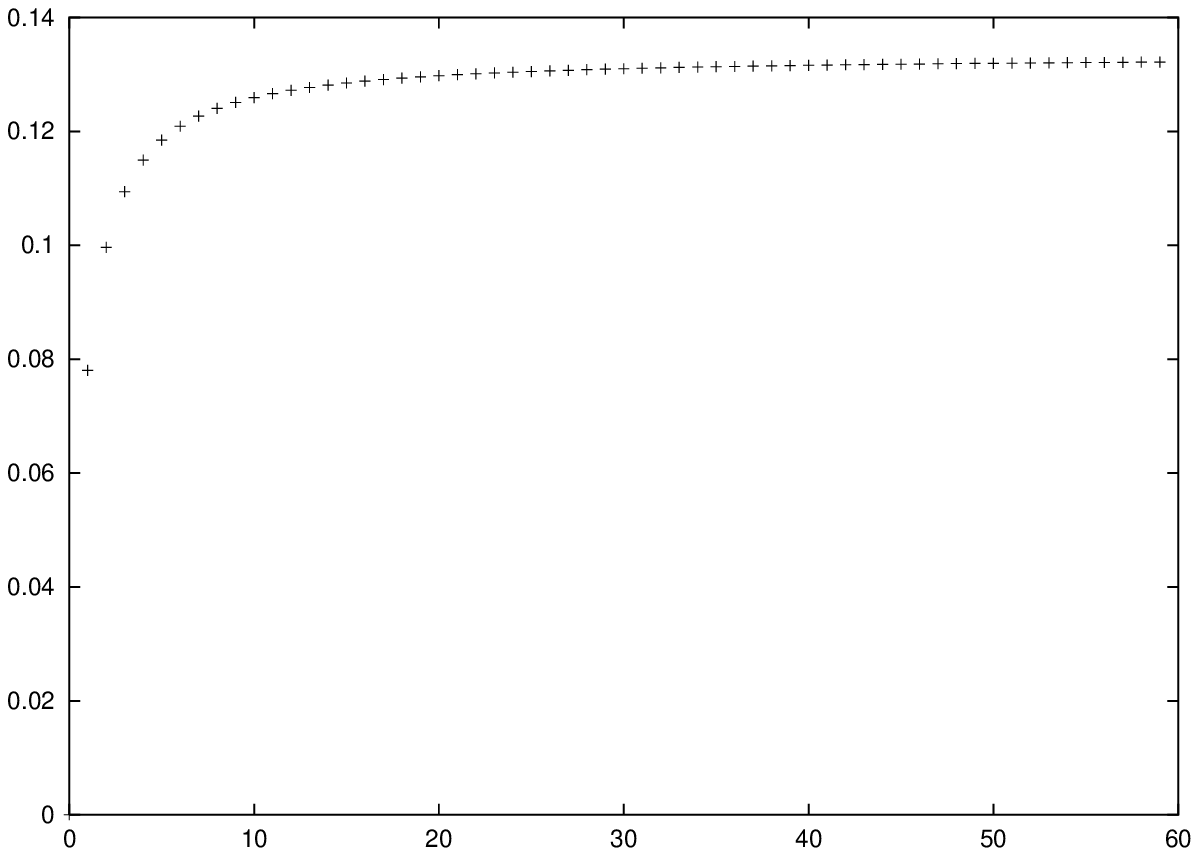}}
b)\resizebox{2.7in}{1.7in}{\includegraphics{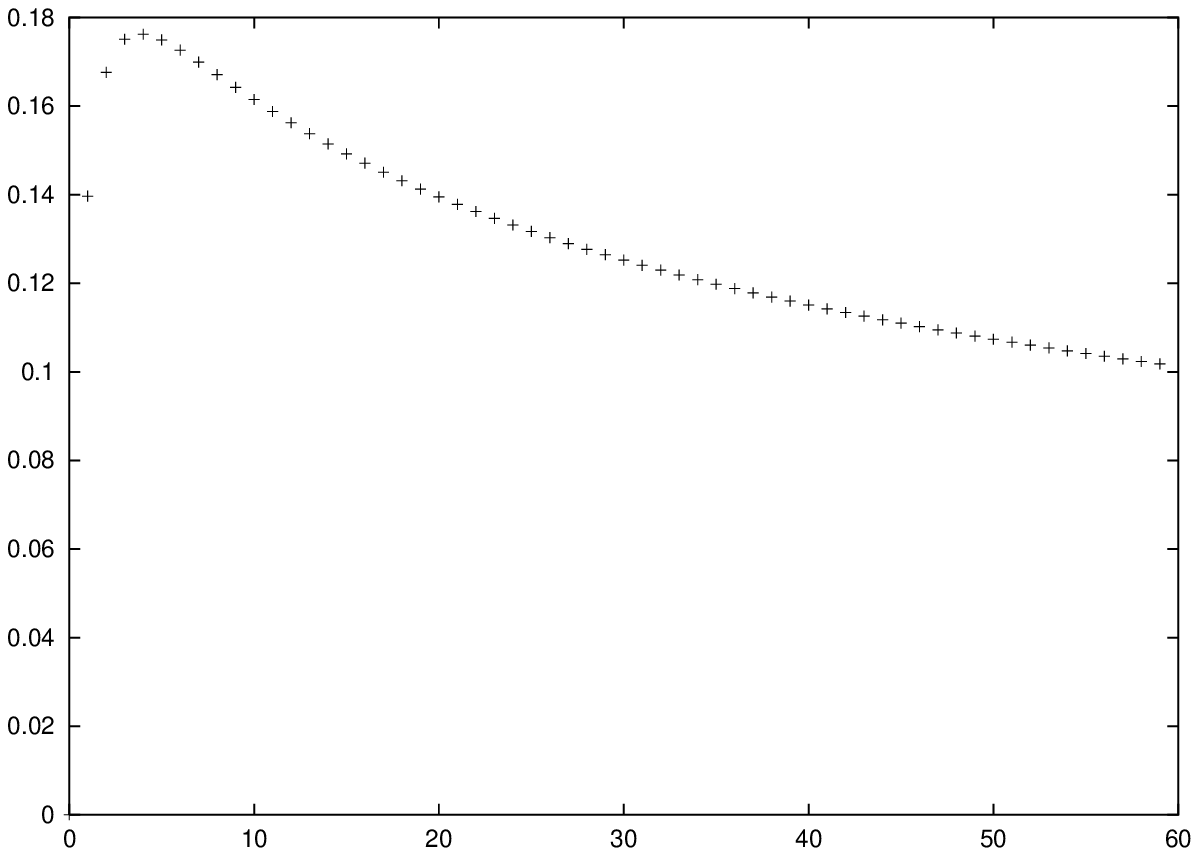}}
\end{center}
Figure 3: a) The norm of $v^L_n$ graphed as a function of $L$ for
identity eq.\ref{matter_e}. b) The inner product $l\cdot v^L$ for
$l_n=\frac{(-1)^n}{2n}$ graphed as a function of $L$ for identity
eq.\ref{matter_e}.
\end{figure}

It is also instructive to see how the midpoint identities converge
in the level truncation scheme. To this end we can calculate the
$L$-component vector $v_n^L$ derived by truncating the right hand
side of one of the equations \ref{matter_e}-\ref{ghost_o} to level
$L$ (by which we mean replace the $M$s by truncated $L\times L$
matrices and the $m$s and $\beta$ by $L$ component vectors).
Presumably as $L\to\infty$ the components of $v^L_n$ should
vanish. In figure 2 we have graphed $|v_m(L)|$ for each identity
eq.\ref{matter_e}-\ref{ghost_o} at levels $L=20,40,...,120$, and
indeed the components seem to fall towards zero, albeit slowly.
Perhaps a little worrisome is the fact that, for fixed $L$,
$v_n^L$ is roughly constant for large values of the mode number
label $n$. This implies that, up to a vanishing factor as
$L\to\infty$, the norm of $v_n^L$ diverges linearly. We have
plotted the norm of $v^L_n$ for identity eq.\ref{matter_e} as a
function of $L$ in figure 3a. As $L$ increases, the norm does not
vanish but seems to approach a constant value, implying that the
components $v^L_n$ fall to zero roughly as $1/\sqrt{L}$. Because
of the distributional nature of the midpoint identities, we should
not necessarily have expected the norm of $v^L$ to vanish at
$L=\infty$, even though its components do. What we really need to
check is that the midpoint identities vanish in a distributional
sense when summed against a vector in $\ell^2$. As a particular
example we can consider the vector $l_n=\frac{(-1)^n}{2n}\in
\ell^2$. We have graphed $l\cdot v^L$ for identity
eq.\ref{matter_e} in figure 3b. For large values of $L$ the inner
product seems to be decreasing towards zero as expected, though
quite slowly---the falloff goes approximately as
$f(L)=(\ln(L)+\gamma)/\sqrt{L}$ ($\gamma=$ Euler's constant).
Using a fitting function $a + bf(L) + cf(L)^2 + df(L)^3$ we can
extrapolate to $L=\infty$ finding $l\cdot v^L\approx -.00088$,
within about half a percent of zero relative to the maximum value
of $l\cdot v^L$. All of this indicates that the vertex does take
the proposed form for well-behaved string fields as the level of
truncation in the theory is increased. However, the convergence is
rather slow, so at low orders in the level expansion in the old
basis one would not expect to immediately see the essential
locality and stability of the theory in lightcone time.

So far, we have not discussed the crucial issue of gauge
invariance of our formalism. The gauge invariance of cubic string
field theory is based on the following axioms\cite{Witten}: 1)
$Q_B$ is nilpotent. 2) The star algebra is associative. 3) The BPZ
inner product is BRST invariant. 4) $Q_B$ acts as a derivation of
the star algebra. Properties 1)-3) can be easily verified in the
tilde basis with a little calculation. Property 4), however, is
notoriously subtle\cite{Gross-Jevicki}, and though we have argued
that the action of the BRST operator on the interaction vertex
should commute with the transformation to the tilde basis, the
BRST operator does involve terms which act quite singularly at the
string midpoint and it would be nice to have an explicit argument
that these operators do not bring additional subtleties. In the
case where we transform only the matter sector to the tilde basis,
the derivation property of $Q_B$ is particularly easy to see and
is worth illustrating. In terms of the interaction vertex, the
derivation property of $Q_B$ is expressed by the formula,
\begin{equation} \sum_{A=1}^3\langle V_3|Q_B^A = 0
\label{derivation}\end{equation} We will show that when the matter
part of the vertex is taken to be as in eq.\ref{our_vertex} and
the BRST operator is expressed as in eq.\ref{BRST}, this property
holds for a large class string fields which are not too singular
at the string midpoint. We write,
\begin{eqnarray}\langle V_3| &=&
\left[\langle\tilde{V}_3|\right]_{k_+=0}\exp(-i\tilde{x}^{+,A}k_+^A)\nonumber\\
Q_B &=& \pi c(\p2)k_+P^+(\p2)-ik_+x^+(\p2)'\pi_b(\p2) +
\tilde{Q}_B|_{k_+=0}
\end{eqnarray} where $\tilde{Q}_B$ and $\langle \tilde{V}_3|$ are
given by their corresponding expressions in the old basis after
the replacement of the oscillators and vacua with their tilde'd
counterparts. Recalling that $c(\p2)$ annihilates the vertex (as
argued before), eq.\ref{derivation} becomes,
\begin{eqnarray} \sum_{A=1}^3\langle V_3|Q_B^A &=&
-i\sum_{A=1}^3\langle V_3|[k_+x^+(\p2)'\pi_b(\p2)]^A +
\left[\sum_{A=1}^3\langle\tilde{V}_3|\tilde{Q}_B^A\right]_{k_+=0}
e^{-i\tilde{x}^{+,B}k_+^B}\  \nonumber\end{eqnarray} Except for
the fact that the oscillators and vacua are tilde'd, the second
term of this equation is exactly the computation of
eq.\ref{derivation} in the old basis, and vanishes for the same
reason, as established for example in ref.\cite{Gross-Jevicki}.
This leaves,
\begin{equation}\sum_{A=1}^3\langle V_3|Q_B^A = -i\sum_{A=1}^3\langle
V_3|[k_+x^+(\p2)'\pi_b(\p2)]^A\label{der_tilde}\end{equation} The
right hand side of this equation now vanishes assuming the overlap
condition,
$$\langle V_3|\pi_b^A(\p2)=0$$ This condition is equivalent to the
identities,\begin{eqnarray}0
&=& \delta_{2m-1}-\tilde{M}_{2m-1,2n-1}\delta_{2n-1}\nonumber\\
0 &=& \tilde{M}^{12}_{2m,2n-1}\delta_{2n-1}\nonumber\end{eqnarray}
where $\delta_{2m-1}$ is as in eq.\ref{delta}. Transforming to the
kappa basis, these identities are
$$0=\frac{\sinh\frac{\pi\k}{2}}{2\cosh\frac{\pi\k}{2}-1}\delta(\k)\
\ \ \ \
0=\left[1-\frac{1}{2\cosh\frac{\pi\k}{2}-1}\right]\delta(\k)$$
Assuming we consider string fields such that the right hand side
of these equations are integrated against test functions which are
smooth at $\k=0$ (i.e. they are not too singular at the midpoint),
$Q_B$ acts as a derivation in the tilde basis and there is no
anomaly in gauge invariance. This is equivalent to saying that
these identities are relations between distributions in the
topological dual of $\ell^2$. One might wonder how the operator
$x^+(\p2)'$, which multiplies $\pi_b(\p2)$ in eq.\ref{der_tilde},
factors into this argument (or similarly, $P^+(\p2)$ which
multiplies $c(\p2)$). It turns out that the action of $x^+(\p2)'$
on the vertex is not as well-defined as that of $\pi_b(\p2)$.
Formally, $x^+(\p2)'$ also annihilates the vertex, but the
``identities'' one derives by imposing $\langle
V_3|x^{+,A}(\p2)'=0$ involve sums like $\sum(-1)^n$ which just
barely diverge. Transforming to the kappa basis, we find that
these identities involve the complex delta function
$\delta(\k-2i)$, which is a more singular distribution than the
real delta function, but nevertheless is well-defined\cite{Erler2}
in the topological dual of the space of test functions which are
analytic on the strip $|\Im(\k)|\leq 2i$\footnote{The sums don't
converge in the mode basis comes because $v_n(\k)$ has a branch
cut singularity at $\k=2i$.}. In mode number language, the
identities $\langle V_3|x^{+,A}(\p2)'=0$ relate distributions in
the topological dual of a half order Sobolev space. Taking extra
care, we might restrict the space of string fields so that the
complex delta function $\delta(\k-2i)$ is always integrated
against an analytic test function, though this is probably
overkill: the fact that $\pi_b(\p2)$ annihilates the vertex is
probably sufficient to ensure the vanishing of the right hand side
of eq.\ref{der_tilde}. Presumably, similar considerations follow
for the case when we transform both matter and ghost sectors to
the tilde basis, though in this case the analysis is more involved
and we will not go into it here.

\section{Derivation of Identities implying Locality}
In this appendix we prove the identities
eq.\ref{matter_e}-\ref{ghost_o} using the spectroscopy of the
Neumann coefficients. We will prove eq.\ref{matter_ek} following
Okuyama\cite{Okuyama2} and leave eq.\ref{matter_ok}-\ref{ghost_ok}
as exercises for the skeptical reader (in fact eq.\ref{ghost_ek}
was already proved in ref.\cite{Okuyama2}). To verify
eq.\ref{matter_ek} we must evaluate the sum eq.\ref{m_k}. For this
we need the explicit form of $m_{2n}$: $$
m_{2m}=-\frac{2}{3}\frac{A_{2n}}{\sqrt{2n}}$$ where the constants
$A_{2n}$ are defined implicitly through the
expansion\cite{Gross-Jevicki},
$$\left(\frac{1+iz}{1-iz}\right)^{1/3} = \exp\left(\frac{2i}{3}
\tan^{-1}z\right)=1+\sum_{n=1}^\infty A_{2n}z^{2n}
+i\sum_{n=1}^\infty A_{2n-1}z^{2n-1}$$ To compute $m(\k)$ insert
the identity in the form,
\begin{eqnarray} 3m(\k) &=& -\frac{1}{\pi i}\oint \frac{dz}{z}\sum_{n=1}^\infty
v_{2n}(\k)z^{-2n}\sum_{m=1}^\infty A_{2m}z^{2m}\nonumber\\
&=& -\frac{1}{\pi i}\oint \frac{dz}{z}\frac{1-\cosh
\k\tan^{-1}z^{-1}}{\k
N(\k)}\exp\left(\frac{2i}{3}\tan^{-1}z\right)
\end{eqnarray} where we used equations \ref{v_gen}. Setting $|z|=1$, $$3m(\k)= -\frac{1}{\pi}\int_{-\pi/2}^{3\pi/2}
d\theta \frac{1-\cosh \k\tan^{-1}e^{-i\theta}}{\k
N(\k)}\exp\left(\frac{2i}{3}\tan^{-1}e^{i\theta}\right)$$ Note
that $\tan^{-1}$ has branch points at $\pm i$ so care must be
taken to ensure we are integrating on the correct branch. This
integral is made much more transparent upon making the
substitutions,
\begin{eqnarray}\tan^{-1}e^{i\theta}&=&\fraction{\pi}{4}+ix\ \ \ \
\ \ \tan^{-1}e^{-i\theta}=\fraction{\pi}{4}-ix\ \ \ \ \ \ d\theta=
\frac{2dx}{\cosh2x}\ \ \ \ \ \theta\in[-\p2,\p2],\ \
x\in[-\infty,\infty]\nonumber\\
\tan^{-1}e^{i\theta}&=&-\fraction{\pi}{4}-ix\ \ \ \
\tan^{-1}e^{-i\theta}= -\fraction{\pi}{4}+ix\ \ \ \ d\theta=
\frac{2dx}{\cosh2x}\ \ \ \ \ \theta\in[\p2,\fraction{3\pi}{2}],\ \
x\in[-\infty,\infty]\nonumber
\end{eqnarray} We have,\begin{eqnarray}3m(\k)&=&
-\frac{2}{\pi}\int_{-\infty}^{\infty} \frac{dx}{\cosh2x}
\frac{1-\cosh \k(\fraction{\pi}{4}-ix)}{\k
N(\k)}\exp\left[\frac{2i}{3}(\fraction{\pi}{4}+ix)\right]\nonumber\\
&\ &\ \ -\frac{2}{\pi}\int_{-\infty}^{\infty} \frac{dx}{\cosh2x}
\frac{1-\cosh \k(\fraction{\pi}{4}-ix)}{\k
N(\k)}\exp\left[\frac{2i}{3}(-\fraction{\pi}{4}-ix)\right]\nonumber\\
&=&-\frac{4}{\pi}\int_{-\infty}^{\infty} \frac{dx}{\cosh2x}
\frac{1-\cosh \k(\fraction{\pi}{4}-ix)}{\k
N(\k)}\cosh\left[\frac{2i}{3}(\fraction{\pi}{4}+ix)\right]\nonumber\\
&=& -\frac{4}{\pi}\frac{1}{\k N(\k)}\int_0^\infty
\frac{dx}{\cosh2x}\left[\sqrt{3}\cosh\fraction{2}{3}x -
\cosh\left(\fraction{\pi
\k}{4}+\fraction{i\pi}{6}\right)\cosh\left(i\k
x+\fraction{2}{3}x\right)\right.\nonumber\\ &\ &\ \ \ \ \ \ \ \
\left.-\cosh\left(\fraction{\pi
\k}{4}-\fraction{i\pi}{6}\right)\cosh\left(i\k
x-\fraction{2}{3}x\right)\right]\nonumber\end{eqnarray} This
integral can be evaluated directly with the help of the formula,
$$\int_0^\infty dx\frac{\cosh ax}{\cosh bx} =
\frac{\pi}{2b}\frac{1}{\cos \frac{\pi a}{2b}} \ \ \ \ \
|\Re(a)|<\Re(b)$$ We find, \begin{eqnarray}3m(\k) &=& -\frac{1}{\k
N(\k)}\left[\frac{\sqrt{3}}{\cos\frac{\pi}{6}}-\frac{\cosh(\frac{\pi\k}{4}
+\frac{i\pi}{6})}{\cos(\frac{\pi}{6}+\frac{i\pi\k}{4})}
-\frac{\cosh(\frac{\pi\k}{4}
-\frac{i\pi}{6})}{\cos(\frac{\pi}{6}-\frac{i\pi\k}{4})}\right]\nonumber\\
&=& -\frac{2}{\k
N(\k)}\frac{\cosh\frac{\pi\k}{2}-1}{1+2\cosh\frac{\pi\k}{2}}\end{eqnarray}
after some algebra. This proves the identity eq.\ref{matter_ek}.

We must now consider the remaining identity, eq.\ref{matter_0},
which is not of the form eq.\ref{matter_ek}-\ref{ghost_ok}. To
prove this formula we must evaluate the sum/integral,
$$3m_{2n}\beta_{2n}=\int_{-\infty}^\infty d\k m(\k)\beta(\k) =
2\int_{-\infty}^\infty d\k
\frac{1-\cosh\frac{\pi\k}{2}}{1+2\cosh\frac{\pi\k}{2}}
\frac{1}{2\k\sinh\frac{\pi\k}{2}}$$ Here we do not need to worry
about the fact that $\beta(\k)$ is proportional to the principal
value distribution since the integrand is smooth at $\k=0$.
Closing the contour in the upper half plane, we find an infinite
succession of poles with corresponding residues,
\begin{eqnarray}\mathrm{Res}(2i(2n-1)) &=& \frac{1}{i\pi}\frac{1}{2n-1}\nonumber\\
\mathrm{Res}(2i(2n-\fraction{4}{3})) &=& -\frac{1}{2\pi
i}\frac{1}{2n-\frac{4}{3}}\nonumber\\
\mathrm{Res}(2i(2n-\fraction{2}{3})) &=& -\frac{1}{2\pi
i}\frac{1}{2n-\frac{2}{3}}\nonumber\end{eqnarray} for
$n=1,2,...,\infty$. We therefore find the sum,
\begin{equation}3m_{2n}\beta_{2n}=2\sum_{n=1}^\infty\left[\frac{2}{2n-1}-\frac{1}{2n-\frac{2}{3}}
-\frac{1}{2n-\frac{4}{3}}\right]\label{sum}\end{equation} This sum
can be evaluated with the help of the formulas,
$$\sum_{n=1}^\infty \frac{1}{n(2n+1)} = 2-\ln 2\ \ \ \ \ \ \
\sum_{n=1}^\infty\frac{1}{n(9n^2-1)} = \frac{3}{2}(\ln 3-1)$$
Writing eq.\ref{sum} in terms of these summations\footnote{Thanks
to N. Mann for doing these sums faster than one of the authors (T.
Erler) could.}, we find
$$3m_{2n}\beta_{2n}=-\ln\frac{27}{16}$$ proving the last identity
eq.\ref{matter_0}.

\section{Useful Formulas} For convenient reference and to set conventions,
here we write some important operators used in the paper. We use
$\alpha'=\half$ so that $\alpha^\mu_0=p^\mu$, and metric signature
$(-,+,+...)$.

\bigskip \noindent Virasoros and $Q_B$:
\begin{eqnarray}L_m &=& \half \sum_{n=-\infty}^\infty
\alpha_{m-n}\cdot\alpha_n\nonumber\\
L_m^{gh}&=& \sum_{n=-\infty}^{\infty}(m-n)b_{m+n}c_{-n}
\nonumber\\ Q_B &=& \sum_{n=-\infty}^\infty :c_n(L_{-n} +\half
L_{-n}^{gh}-\delta_{n0}): \nonumber
\end{eqnarray}

\noindent String coordinates: \begin{eqnarray} x^\mu(\sigma) &=&
x^\mu+i\sum_{n=1}^\infty
\frac{1}{2n}(\alpha^\mu_n-\alpha^\mu_{-n})\cos n\sigma \nonumber\\
\pi P^\mu(\sigma) &=& p^\mu+\sum_{n=1}^\infty
(\alpha^\mu_n+\alpha^\mu_{-n})\cos
n\sigma\ \ \ \nonumber\\
b(\sigma)&=&-\sum_{n=1}^\infty (b_n-b_{-n})\sin
n\sigma \nonumber\\
\pi\cdot\pi_c(\sigma) &=& b_0 + \sum_{n=1}^\infty (b_n+b_{-n})\cos
n\sigma \nonumber\\
c(\sigma)&=&c_0+\sum_{n=1}^\infty (c_n+c_{-n})\cos n\sigma \nonumber\\
\pi\cdot\pi_b(\sigma) &=& \sum_{n=1}^\infty (c_n-c_{-n})\sin
n\sigma. \nonumber\end{eqnarray}

\noindent Midpoint coordinates:
\begin{eqnarray}x^\mu(\p2) &=& x^\mu + i\sum_{n=1}^\infty
\fraction{(-1)^n}{2n}(\alpha^\mu_{2n}-\alpha^\mu_{-2n})\nonumber\\
\pi \cdot P^\mu(\p2) &=& p^\mu + \sum_{n=1}^\infty
(-1)^n(\alpha^\mu_{2n}+\alpha^\mu_{-2n})\nonumber\\
x^\mu(\p2)' &=& i\sum_{n=1}^\infty
(-1)^n(\alpha^\mu_{2n-1}-\alpha^\mu_{-2n+1})\nonumber\\
\pi\cdot P(\p2)' &=& \sum_{n=1}^\infty
(2n-1)(-1)^n(\alpha^\mu_{2n-1}+\alpha^\mu_{-2n+1})\nonumber\\
b(\p2) &=& \sum_{n=1}^\infty(-1)^n(b_{2n-1}-b_{-2n+1})\nonumber\\
b(\p2)' &=& -\sum_{n=1}^\infty 2n(-1)^n(b_{2n}-b_{-2n})\nonumber\\
\pi\cdot\pi_c(\p2) &=& b_0 +
\sum_{n=1}^\infty(-1)^n(b_{2n}+b_{-2n})\nonumber\\
\pi\cdot\pi_c(\p2)' &=&
\sum_{n=1}^\infty(2n-1)(-1)^n(b_{2n}+b_{-2n})\nonumber\\
c(\p2) &=& c_0 + \sum_{n=1}^\infty
(-1)^n(c_{2n}+c_{-2n})\nonumber\\
c(\p2)' &=& \sum_{n=1}^\infty
(2n-1)(-1)^n(c_{2n-1}+c_{-2n+1})\nonumber\\
\pi\cdot\pi_b(\p2) &=&
-\sum_{n=1}^\infty(-1)^n(c_{2n-1}-c_{-2n+1})\nonumber\\
\pi\cdot\pi_b(\p2)' &=& \sum_{n=1}^\infty
2n(-1)^n(c_{2n}-c_{-2n})\nonumber\end{eqnarray} We have written
these in terms of oscillators in the usual basis, but whenever we
refer to them in the paper (with the exception of eq.\ref{pic})
they happen to take the same form in the tilde basis.
\end{appendix}

\end{document}